\documentclass[%
 reprint,
 superscriptaddress,
 nofootinbib,
 amsmath,amssymb,
 aps,
 prx,
]{revtex4-2}

\usepackage[hidelinks]{hyperref}
\usepackage{graphicx}
\usepackage{dcolumn}
\usepackage{bm}
\usepackage{booktabs}
\usepackage{amsmath,amssymb,amsbsy,amstext, amsthm, amsfonts}
\usepackage{color}
\usepackage{slashed}
\usepackage[dvipsnames]{xcolor}
\usepackage{dsfont}
\usepackage{verbatim}
\usepackage[utf8]{inputenc} 
\usepackage{ragged2e}
\usepackage{mathtools}
\usepackage{xfrac}
\usepackage{blkarray}

\def\Vhrulefill{\leavevmode\leaders\hrule height 0.7ex depth \dimexpr0.4pt-0.7ex\hfill\kern0pt}

\begin{document}

\title{Higher Flavor Symmetries in the Standard Model}

\author{Clay C\'{o}rdova}
\affiliation{Enrico Fermi Institute, University of Chicago}
\affiliation{Kadanoff Center for Theoretical Physics, University of Chicago}
\author{Seth Koren}
\affiliation{Enrico Fermi Institute, University of Chicago}

\date{\today}

\begin{abstract}
We initiate the study of the generalized global flavor symmetries of the Standard Model.
The presence of nonzero triangle diagrams between the $U(3)^5$ flavor currents and the $U(1)_Y$ hypercharge current intertwines them in the form of a higher-group which mixes the zero-form flavor symmetries with the one-form magnetic hypercharge symmetry.  
This higher symmetry structure greatly restricts the possible flavor symmetries that may remain unbroken in any ultraviolet completion that includes magnetic monopoles.  
In the context of unification, this implies tight constraints on the combinations of fermion species which may be joined into multiplets.
Three of four elementary possibilities are reflected in the classic unification models of Georgi-Glashow, $SO(10)$, and Pati-Salam.  
The final pattern is realized non-trivially in trinification, which exhibits the sense in which Standard Model Yukawa couplings which violate these flavor symmetries may be thought of as spurions of the higher-group.
Such modifications of the ultraviolet flavor symmetries are possible only if new vector-like matter is introduced with masses suppressed from the unification scale by the Yukawa couplings.  
\end{abstract}

\maketitle

\makeatletter
\def\l@subsubsection#1#2{}
\makeatother


\section{Introduction}

In this work we begin an exploration of the Standard Model's flavor symmetries incorporating the novel effects of higher symmetries \cite{Gaiotto:2014kfa}.  While the flavor symmetries are at best approximate, it is an experimental fact that flavor-violating processes are anything but generic. We have long appreciated that understanding precisely how the flavor symmetries are broken by the Yukawa matrices of the Standard Model yields powerful information about the structure of flavor-changing effects in the infrared as in the celebrated GIM mechanism \cite{Glashow:1970gm}. Similarly, systematically organizing the effects of flavor symmetry-breaking continues to guide work on ultraviolet physics beyond the Standard Model \cite{DAmbrosio:2002vsn}.

\begin{figure}[h]
    \centering
        {\includegraphics[clip, trim=0.0cm 0.0cm 0.0cm 0.0cm, width=0.35\textwidth]{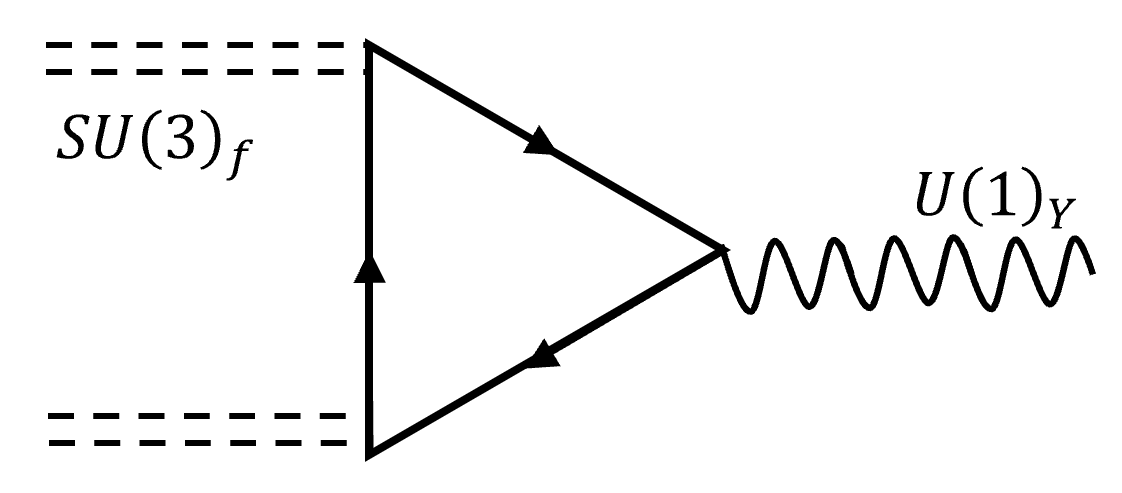}}
    \caption{A triangle diagram between the gauged hypercharge current and two global flavor currents leads to the Standard Model two-group structure we will focus on below.}
    \label{fig:gGG}
\end{figure}

Following this tradition, below we will analyze the symmetries of the Standard Model in the limit of vanishing Yukawa couplings.  Our goal is to elucidate the interplay between these familiar (zero-form) flavor symmetries, and the higher (one-form) magnetic hypercharge symmetry which is present in the Standard Model because there are no dynamical magnetic monopoles. In particular, we show that as a consequence of the triangle diagrams illustrated in Figure \ref{fig:gGG}, the nonabelian flavor symmetries and the magnetic one-form symmetry are unified into a composite structure known as a higher-group \cite{Kapustin:2013uxa, Tachikawa:2017gyf, Cordova:2018cvg,Benini:2018reh}. 
Meanwhile, the structure of the abelian chiral symmetries of the Standard Model is even more delicate, forming a non-invertible algebraic symmetry structure of the type discussed in \cite{Choi:2022jqy, Cordova:2022ieu}. Other recent work toward higher symmetries in particle physics models includes \cite{Hidaka:2020izy,Choi:2022rfe,Cordova:2022fhg, Hsin:2022heo,Brennan:2020ehu,Choi:2022fgx,Wang:2022fzc,Hinterbichler:2022agn,Cordova:2022rer,Anber:2021iip,Fan:2021ntg,Hong:2020bvq,Brennan:2023xyz,Heidenreich:2020pkc,McNamara:2022lrw}, with \cite{Wang:2018jkc,Wan:2019gqr,Davighi:2019rcd,Wang:2020mra,Wang:2020xyo,Anber:2021upc,Wang:2021ayd,Wang:2021vki,Wang:2021hob,Davighi:2022icj,Davighi:2020uab} touching on aspects of the Standard Model structure.

We utilize these ideas to constrain the possible patterns of flavor symmetries in any ultraviolet model with magnetic monopoles.  The presence of such particles implies that at short distances magnetic charge can be screened, and hence the one-form magnetic symmetry is broken.   This is the electromagnetic dual of usual screening of test charges by virtual pairs of electrically charged particles.  Following \cite{Cordova:2018cvg,Benini:2018reh}, we argue that with only the Standard Model matter content, any flavor symmetry that participates non-trivially in a higher algebraic structure with this magnetic symmetry is also necessarily explicitly broken in the ultraviolet (even in the absence of Yukawa couplings).  Letting $E_{\text{flavor}}$ denote the energy scale where a given flavor symmetry emerges, and $E_{\text{magnetic}}$ the symmetry emergence scale of the one-form magnetic symmetry, we obtain universal inequalities:\footnote{For related inequalities in models of axions see \cite{Brennan:2020ehu,Choi:2022fgx}.}
\begin{equation}\label{ineq}
    E_{\text{flavor}} \lesssim  E_{\text{magnetic}}~.
\end{equation}

In the context of models of unification with a weakly coupled ultraviolet gauge group, the scale $E_{\text{magnetic}}$ is the higgsing scale where $U(1)_{Y}$ emerges \cite{Cordova:2022rer}.
The most basic application of \eqref{ineq} therein is to the `flavor symmetries' of a single generation of Standard Model fields: the independent $U(1)$ rotations of different species in the infrared which descend from the same ultraviolet multiplet necessarily \textit{emerge} below the scale of unification. In the ultraviolet it is only simultaneous $U(1)$ rotations of the entire multiplet which are good symmetries, whereas any orthogonal subgroup of the infrared flavor symmetries is broken by the gauge theory in which hypercharge is embedded. 
As it is only flavor symmetries which participate in the higher algebraic structure which are subject to \eqref{ineq}, understanding precisely the Standard Model symmetries then constrains the allowed patterns of unification for the Standard Model fermions.

With the Yukawa couplings of the Standard Model turned off, we will find robustly only four possible unification patterns compatible with the Standard Model gauge structure in the infrared.
However, to fully see the role of the higher structure in controlling the ultraviolet of the Standard Model with non-zero Yukawa couplings, we must understand the impact of the higher flavor symmetries being approximate.

In the following section we will explain the sense in which Yukawas which violate the zero-form flavor symmetries may be seen as \textit{spurions} for the modification of the higher flavor structure.
We illustrate in detail how this higher-symmetry violation appears in both a toy model and in a well-motivated unification scheme for physics beyond the Standard Model. This potentially opens the door to future applications of higher symmetry structures in particle physics, where zero-form symmetries of interest are often only approximate.

In particular, we will see that the structure of the Standard Model approximate higher flavor-hypercharge symmetry may be modified by adding matter which is vector-like under the Standard Model gauge group (and hence potentially massive), but chiral under the infrared flavor symmetry.  Hence, the introduction of such matter allows the realization of more general unification patterns.

\begin{figure}[h]
    \centering
        {\includegraphics[clip, trim=0.0cm 0.0cm 0.0cm 0.0cm, width=0.5\textwidth]{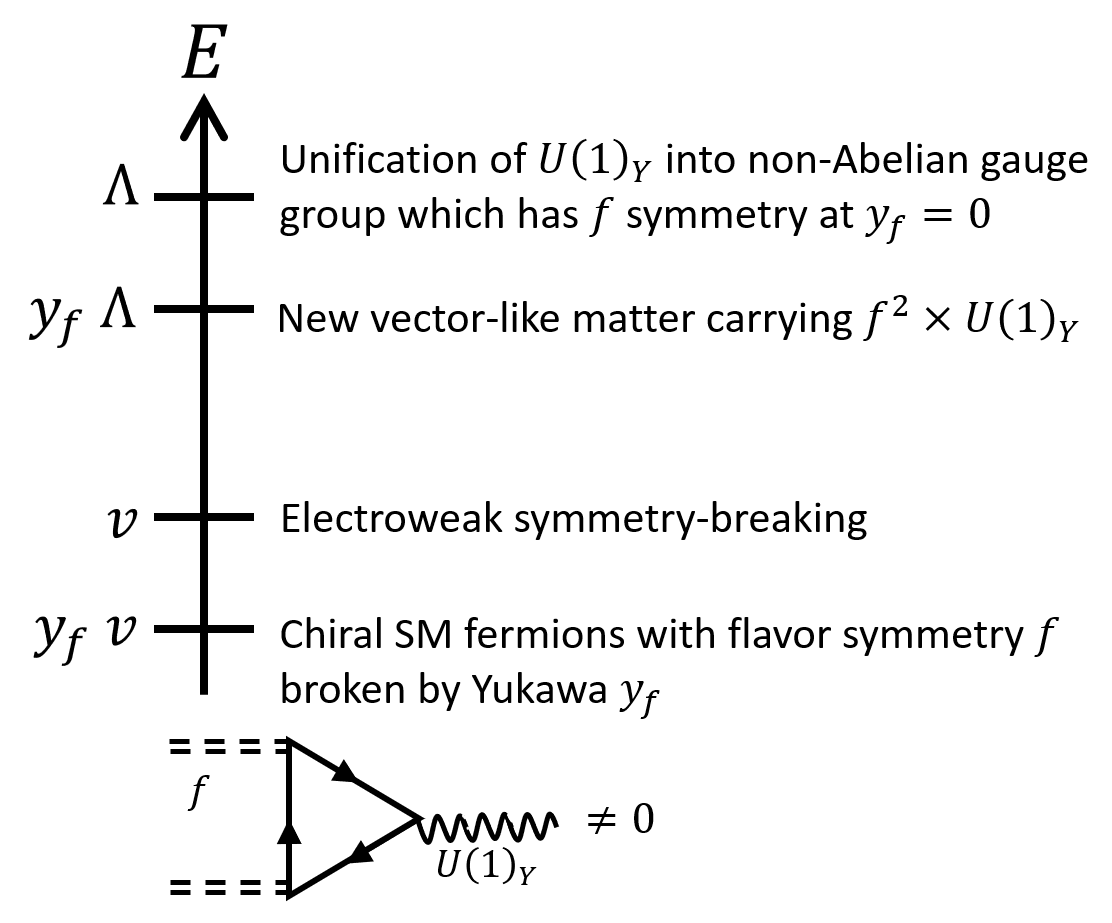}}
    \caption{
    Symmetry-breaking scales in the Standard Model with approximate flavor symmetry $f$ broken by a Yukawa $y_f$.
    To the extent that the flavor symmetry is approximate, its higher-group structure may be modified by vector-like matter which lies parametrically below the unification scale.
    }
    \label{fig:scales}
\end{figure}

However, below we show that the presence of the approximate higher flavor symmetry implies that the mass scale of the new vector-like matter necessarily vanishes in the limit of zero Standard Model Yukawa couplings when the flavor symmetry is restored.  Viewed from the ultraviolet, this means that the same interactions that generate the Standard Model Yukawa couplings also necessarily generate the masses for this new matter. As such, realizing ultraviolet flavor symmetry patterns beyond those classified at zero Yukawas with only the Standard Model spectrum requires new matter whose masses are suppressed from the unification scale by the Yukawa couplings.  Thus, to the extent that the flavor symmetries are approximately valid, i.e.\ for small Yukawa couplings, vector-like matter modifying our conclusions is parametrically lighter than the unification scale. (See Figure \ref{fig:scales}.)

Thus, while an exact flavor symmetry allows one to infer exactly the flavor-chiral matter at the unification scale $\Lambda$ from the far infrared, an approximate flavor symmetry broken by a coupling $y$ implies one must understand the spectrum from the infrared up to a scale $\sim y \Lambda$ before one has nailed the flavor-chiral matter present at $\Lambda$. 
Nevertheless, the higher flavor structure is restrictive, and even with approximate flavor symmetries still tightly constrains the overall patterns of unification of the Standard Model fermions.

In the following, we briefly review relevant aspects of generalized global symmetries.  We then present a simple analysis in a toy model illustrating the essential logic.  Finally, we apply these ideas to the Standard Model and illustrate the conclusion in various well known models where hypercharge emerges from a nonabelian group.  

\section{Generalized Symmetries and Emergence Scales}

Let us review the basic concepts of generalized global symmetries which will feature in our analysis.  

Standard continuous global symmetries are encoded in quantum field theory by the presence of currents $J_{\mu}(x)$ which are conserved at separated points in correlation functions.  Such currents imply familiar selection rules on scattering amplitudes.  In modern parlance, these familiar currents generate \textit{zero-form} symmetries. They act on local, zero-dimensional (point-like) operators such as the gauge invariant composites of fields in a Lagrangian, or equivalently on the  particles created by such operators.  

Continuous higher-form global symmetry is similarly characterized by conserved currents, but as compared to the case above these currents carry additional spacetime indices \cite{Gaiotto:2014kfa}. In our application we will be focused in particular on \emph{one-form} global symmetries.  The associated currents are then local operators $J_{\mu\nu}(x)$ which are antisymmetric in their spacetime indices and obey a conservation equation $\partial^\mu J_{\mu\nu}(x) = 0$.  We note also that unlike ordinary symmetries which may be abelian or nonabelian, one-form symmetries are necessarily abelian.

The paradigmatic example of one-form symmetry occurs in free Maxwell theory, i.e.\ $U(1)_{g}$ gauge theory without charged matter.  In this simple theory, the equation of motion and the Bianchi identity may be viewed as conservation equations for two distinct one-form symmetry currents conventionally referred to as electric ($E$), and magnetic ($M$):
\begin{equation}
    J^E_{\mu\nu} \sim\frac{1}{e^{2}}F_{\mu\nu}~, \hspace{.2in} 
    J^M_{\mu\nu} \sim\varepsilon_{\mu\nu\rho \sigma}F^{\rho \sigma}.
    \label{eq:one-form_current}
\end{equation}
Unlike ordinary symmetries, these one-form symmetries do not act on local operators.  Instead their charges are computed by surface integrals, the simplest of which are localized in a fixed time slice.  In this case the charges are reduced to familiar Gaussian surface integrals:
\begin{equation}\label{1formcharges}
    Q^{E}(\Sigma)\sim\int_{\Sigma} \mathbf{E}\cdot \mathbf{dS}~, \hspace{.2in}  Q^{M}(\Sigma)\sim\int_{\Sigma} \mathbf{B}\cdot \mathbf{dS}~.
\end{equation}
The field configurations which activate these one-form symmetry charges are those of infinitely massive source particles carrying definite electric and magnetic charges.  Viewed in spacetime such sources trace out worldlines defining  line operators in the field theory which are thus the natural objects charged under one-form symmetry.  In Maxwell theory, these are electrically charged Wilson lines, and magnetically charged 't Hooft lines.   

Let us now turn to symmetry breaking.  Since all local operators are neutral under one-form symmetries we cannot violate the conservation of \eqref{eq:one-form_current} by adding local operators to the Lagrangian.\footnote{More technically, in the presence of higher derivative corrections, the electric one-form current must be modified but it is always possible to recast the resulting equations of motion as current conservation conditions \cite{Cordova:2022rer}.}  Instead the symmetries are violated when the higher-form current conservation is fundamentally broken by altering the degrees of freedom.  In familiar weakly-coupled gauge theories this leads to the following possibilities:
\begin{itemize}
    \item The electric one-form symmetry current $\frac{1}{e^{2}}F_{\mu\nu}$ is violated when there is finite mass electrically charged matter.  This is because such matter fields may screen the electric charges of Wilson lines.\footnote{If the electric charges of the matter fields are all divisible by a common integer multiple $q>1$, then the one-form symmetry is broken instead to a non-trivial discrete subgroup. This phenomenon will not play a role in the following.}
    \item The magnetic one-form symmetry $\varepsilon_{\mu\nu\rho \sigma}F^{\rho \sigma}$ can similarly be broken by the presence of finite mass magnetic monopoles, or relatedly by realizing the abelian gauge field via higgsing from a semi-simple gauge group (i.e.\ a gauge group with no $U(1)$ factors) no matter how heavy its physical monopoles are.  In the following, we let $\Phi$ denote such a generic higgs field and $\Lambda$ the scale of its expectation value.  
    Above the scale $\Lambda$ it is clear that there are no gauge invariant two-form operators satisfying the Bianchi identity and hence the magnetic one-form symmetry is violated.
\end{itemize}
In practice, we will be interested in abelian gauge theories with charged matter. In this case, the electric one-form symmetry is broken, but the magnetic one-form symmetry remains.  In particular, in the context of the Standard Model, the abelian hypercharge gauge group generates a magnetic one-form global symmetry.  From the discussion above, in UV completions of the Standard Model this one-form symmetry is typically accidental, and is broken at a unification scale $\Lambda$ where $U(1)_{Y}$ is embedded into a nonabelian group.     

\subsection{Higher Groups and the Emergence Theorem }\label{sec:emergence}

In general, we may consider models with both zero-form and one-form symmetry, and it is natural to ask about their possible interplay.  Two broad phenomena have been recently investigated.  

 \emph{Non-Invertible Symmetry} \cite{Choi:2021kmx, Kaidi:2021xfk}: This structure arises in particular when there are non-trivial triangle anomaly coefficients with vertices gauge-gauge-flavor. (Note that such a non-zero triangle is possible only for abelian flavor symmetries.)  
    When the gauge vertices are nonabelian this anomaly breaks the flavor symmetry in question through instanton processes.  When the gauge vertices are abelian the symmetry is not violated, but instead its charges form a non-trivial algebra with those defined in \eqref{1formcharges} characterizing the one-form symmetry \cite{Choi:2022jqy, Cordova:2022ieu}.  
    
    Such non-invertible symmetries arise in the limit of the Standard Model with vanishing Yukawa couplings where the flavor symmetry $U(1)_{\bar{u}-\bar{d}+\bar{e}}$ acting by phase rotations on the indicated Standard-Model fields (see Table \ref{tab:charges}) is intertwined with the one-form magnetic hypercharge symmetry.\footnote{As observed in \cite{Choi:2022jqy}, this non-invertible symmetry may be applied to deduce the coupling of the neutral pion to photons.} More generally, such non-invertible symmetries are a generic feature of models of $Z'$ gauge bosons \cite{Cordova:2022fhg}.  These symmetries, while fascinating, are not our focus in the present work.
    
 \emph{Two-Group Symmetry} \cite{Kapustin:2013uxa, Cordova:2018cvg, Benini:2018reh}: This structure arises when there are non-trivial triangle anomaly coefficients with vertices gauge-flavor-flavor.  (Note that such a non-zero triangle is possible only for abelian gauge vertices.) In general, we denote by $\kappa\in \mathbb{Z}$ the quantized triangle coefficient. For an abelian gauge theory with left-handed Weyl fermions of gauge charge $q_{g}$ and $U(1)_{f}$ flavor charges $q_{f}$ we have
 \begin{equation}\label{weylsumform}
     \kappa = \sum_{\text{Weyls}} q_{g}q_{f}^{2}\in \mathbb{Z}~.
 \end{equation}
 More generally for a nonabelian flavor symmetry, the $q_{f}^{2}$ term is replaced by a quadratic Casimir.  We also note that such triangle coefficients are insensitive to the sign of $q_{f}$ (in the nonabelian case, this means they are invariant under complex conjugation, i.e.\ changing fundamentals to antifundamentals).

 When $\kappa \neq0$ one says there is two-group global symmetry.  In this case the flavor symmetry is still present in the theory, i.e.\ the current $J_{\mu}(x)$ is conserved at separated points in correlation functions and its associated conserved charges organize the Hilbert space into representations.  However, there is a non-trivial current algebra involving the ordinary flavor currents and the magnetic one-form symmetry current \cite{Cordova:2018cvg}.  For instance, the magnetic one-form symmetry current appears as a contact term in the operator product expansion of the ordinary currents $J_{\mu}(x)$:
\begin{equation}\label{eq:oneOneTwo}
    \partial^{\mu}J_{\mu}(x) \cdot J_{\nu}(y)\sim \kappa \  \partial^{\lambda}\delta(x-y)\varepsilon_{\lambda \nu \rho \sigma}F^{\rho \sigma}(y)~.
\end{equation}
This equation is reminiscent of anomalous Ward-identities, but with the crucial difference that the right-hand side of the above is a non-trivial operator in the theory.\footnote{In the notation used in \cite{Cordova:2018cvg}, the one-form symmetry group is  $U(1)^{(1)}$, the zero-form symmetry group is $f^{(0)}$ and the two-group together with its structure constant $\kappa$ is written as $f^{(0)}\times_{\hat{\kappa}} U(1)^{(1)}$ with $\hat{\kappa}=-\kappa/2.$ }  

Crucial to us is the following observation of \cite{Cordova:2018cvg} which is manifest in \eqref{eq:oneOneTwo}: when $\kappa$ is non-vanishing the operator algebra containing the ordinary flavor symmetry currents also contains the magnetic one-form symmetry current.  Thus, whenever the ordinary flavor symmetries are present, and $\kappa$ is non-zero, the magnetic one-form symmetry current (i.e.\ the abelian gauge field strength) is also necessarily part of the local operator spectrum.   

In this way the two-group current algebra is analogous to a nonabelian symmetry though formed out of currents of different spin, and where $\kappa$ plays the role of the structure constant.  Moreover, as in that context the coefficient $\kappa \in \mathbb{Z}$ is quantized and hence cannot change continuously under renormalization group flow, nor any other deformation of the theory that preserves the flavor symmetry.      

These observations lead directly to the two-group emergence theorem established in \cite{Cordova:2018cvg} and further applied in \cite{Benini:2018reh,Brennan:2020ehu,Choi:2022fgx} which directly constrains allowed patterns of two-group symmetry breaking.  In a two-group with non-zero $\kappa$, the zero-form currents do not close under the operator product.  Therefore, for non-zero $\kappa$ breaking the one-form symmetry without breaking the zero-form symmetry is impossible. With an eye towards later applications we state a version of this theorem as follows:

\emph{Consider any effective field theory with an abelian gauge field $U(1)_{g}$ and a flavor symmetry $f$ with a non-zero flavor-flavor-gauge triangle coefficient $\kappa$.  Then any ultraviolet where $U(1)_{g}$ emerges via higgsing from a semi-simple group necessarily breaks the flavor symmetry $f$ at energy scales no greater than the higgsing scale $\Lambda.$}

More bluntly, unification of $U(1)_{g}$ into a nonabelian group where $U(1)_{g}$ is realized as a traceless generator is incompatible with the flavor symmetry with non-zero flavor-flavor-gauge  triangle coefficient $\kappa$.  This establishes the inequality of scales stated in \eqref{ineq}. Physically, if the flavor symmetry currents can turn on a one-form symmetry current, these flavor symmetries cannot be present above the scale at which the one-form symmetry is broken by the nonabelian gauge dynamics.

Thus far we have reviewed the analysis of \cite{Cordova:2018cvg} which utilized exact flavor symmetry.  In our application we are interested in generalizing these ideas to situations where the flavor symmetry is broken explicitly by interactions in the Lagrangian, as in the Standard Model.  We can take into account flavor symmetry breaking effects using the following essential logic.

 Consider any effective field theory $\mathcal{T}_{\text{IR}}$ with an abelian gauge field $U(1)_{g}$. We let $y$ denote a dimensionless coupling such that when $y\rightarrow 0$ a flavor symmetry $f$ is restored.  To indicate the dependence on $y$, we refer to the theory as $\mathcal{T}_{\text{IR}}(y).$ In this situation the parameter $y$ is a spurion for the flavor symmetry, though presently we will see that it is often fruitful to view it as a spurion of a two-group.

 The effective field theory in question is the low energy limit of an ultraviolet theory which is also necessarily a function of the parameter $y$.  We denote this theory as $\mathcal{T}_{\text{UV}}(y)$.    Hence under the renormalization group:
    \begin{equation}
        \text{RG}: \mathcal{T}_{\text{UV}}(y) \longrightarrow \mathcal{T}_{\text{IR}}(y)~.
    \end{equation}
We assume that in $\mathcal{T}_{\text{UV}}(y)$ the magnetic one-form symmetry arising from $U(1)_{g}$ is broken.  (For example this occurs if the abelian gauge group arises from higgsing a nonabelian group with no $U(1)$ factors.)  By $\mathcal{T}_{\text{IR}}(y)$ we thus mean the effective field theory describing all degrees of freedom below the one-form symmetry emergence scale (e.g.\ the higgsing scale) $\Lambda$.

 We ask whether in the limit $y\rightarrow 0$ the UV theory $\mathcal{T}_{\text{UV}}(y)|_{y=0}$ can also enjoy the flavor symmetry $f$. This is in principle possible since the IR theory $\mathcal{T}_{\text{IR}}(y)|_{y=0}$ indeed has the flavor symmetry.  According to the two-group emergence theorem of \cite{Cordova:2018cvg} referenced above, this is only possible if in the theory $\mathcal{T}_{\text{IR}}(y)|_{y=0}$, the flavor-flavor-gauge triangle coefficient $\kappa$ vanishes.
    
 Note that the parameter $y$ may control the masses of many fields.  The flavor-flavor-gauge triangle coefficient receives contributions from all fields which become massless in the limit $y\rightarrow 0$.  These include for instance massless fermions, and also Goldstone bosons which may carry Wess-Zumino terms.\footnote{We thank K. Harigaya and L.T. Wang for discussions related to Froggatt-Nielsen scenarios \cite{Froggatt:1978nt}.}  When $y\neq 0$ these fields contributing to $\kappa$ may have a variety of different mass scales, but all must necessarily vanish when $y\rightarrow 0$.   At non-zero $y$ we can thus partition the degrees of freedom contributing to $\kappa$ as follows:
 \begin{itemize}
     \item Massless degrees of freedom in $\mathcal{T}_{\text{IR}}(y)$.  We denote their contribution to $\kappa$ as $\kappa_{0}.$
     \item A sequence of massive degrees of freedom whose masses are $y^{\alpha_{i}}m_{i},$ where $\alpha_{i}>0$ and $m_{i}$ is a mass scale.   We denote each of their contributions as $\kappa_{i}.$    
 \end{itemize}
     By definition we then have the following sum rule:
     \begin{equation}\label{sumrule}
         \kappa=\kappa_{0}+\sum_{i}\kappa_{i}~,
     \end{equation}
and the ultraviolet $\mathcal{T}_{\text{UV}}(y)$ can only restore the flavor symmetry when $y=0$ if the sum above vanishes.  

As a corollary of these ideas we note that if $\kappa_{0}\neq 0$ and no additional degrees of freedom appear below the scale $\Lambda$, then $\mathcal{T}_{\text{UV}}(y)$ cannot enjoy the flavor symmetry $f$ even in the limit $y\rightarrow 0$.  Conversely if $\mathcal{T}_{\text{UV}}(y)|_{y=0}$ does have the flavor symmetry $f$ and $\kappa_{0}\neq 0 $ then there are necessarily new degrees of freedom below the UV scale $\Lambda$ which contribute non-zero $\kappa_{i}$ in \eqref{sumrule} and whose masses are parametrically suppressed by a positive power of $y$. In particular, in this scenario, the same ultraviolet coupling that violates the flavor symmetry also generates the masses for the additional degrees of freedom.

In practice, we can apply this argument to constrain the possible maximal flavor symmetry of UV completions $\mathcal{T}_{\text{UV}}(y)$ of a given IR theory $\mathcal{T}_{\text{IR}}(y)$. In a gauge theory, this maximal flavor symmetry is determined by the multiplet structure of the matter and thus constrains unification patterns.  We illustrate this below in a toy example before applying it directly to the Standard Model.

\subsection{A Toy Model}

{\setlength{\tabcolsep}{0.6 em}
\renewcommand{\arraystretch}{1.3}
\begin{table}[h!]\centering
\large
\begin{tabular}{|c|c|c|c|}  \hline
 & $U(1)_g$ & $SU(N_f)_L$ & $SU(N_f)_R$  \\ \hline
$\chi^+$ & $+1$ & $\Box$ & -- \\
$\chi^-$ & $-1$ & -- & $\Box$ \\ \hline
\end{tabular}\caption{Infrared vector-like abelian gauge theory.}\label{tab:toyIR}
\end{table}}

As a warm-up, we wish to study a simple toy model  which, while not containing the richness of the Standard Model, allows us to cleanly exhibit the basic implications of the two-group structure and the sense in which flavor-violating couplings are spurions for the higher-group. 
The simplicity means that some of the conclusions at which we arrive one could have easily guessed, but the conclusions below when applied to the Standard Model will be less obvious.  

We consider a vector-like infrared $U(1)_g$ abelian gauge theory with $N_f$ pairs of massless fermions $\chi^+_i, \chi^-_i$ which enjoy separate nonabelian flavor symmetries $SU(N_f)_{L,R}$, as displayed in Table \ref{tab:toyIR}. Both of these flavor symmetries have nonzero gauge-global-global anomalies with $U(1)_g$, listed in Table \ref{tab:toyAnom}, so as first analyzed in \cite{Cordova:2018cvg}, participate in a two-group symmetry. 

{\setlength{\tabcolsep}{0.6 em}
\renewcommand{\arraystretch}{1.3}
\begin{table}[h!]\centering
\large
\begin{tabular}{|c|c|}  \hline
\text{Flavor}$^2$ & $U(1)_g$ \\ \hline
$SU(N_f)_L^2$ & $+ 1$ \\ 
$SU(N_f)_R^2$ & $- 1$ \\ \hline
\end{tabular}\caption{gauge-global-global anomalies of the toy model.}\label{tab:toyAnom}
\end{table}}

Now let us try to `UV complete' this $U(1)_g$ gauge theory into a nonabelian $SU(2)_g$ gauge theory. 
Following the emergence theorem above, the separate flavor symmetries of $\chi^+$ and $\chi^-$ cannot be good UV flavor symmetries.
However, the diagonal subgroup thereof in which both $\chi^+$ and $\chi^-$ transform as fundamentals is not part of the two-group, so may be a good UV symmetry. 

Indeed, this guides us to the obvious UV completion in which $(\chi^+_i, \chi^-_i)$ are embedded in a single $SU(2)_g$ fundamental as in Table \ref{tab:toyUVExact}.\footnote{To avoid the Witten anomaly \cite{Witten:1982fp}, we assume that $N_{f}$ is even.} 
This UV theory has a single $SU(N_f)$ flavor symmetry, and upon higgsing $SU(2)_g \rightarrow U(1)_g$ by a triplet scalar $\Phi$ this returns to the diagonal flavor symmetry of the infrared. 
The separate $SU(N_f)_L, SU(N_f)_R$ symmetries emerge below the scale $\langle \Phi \rangle = \Lambda$ at which the UV gauge group breaks and the $U(1)_g^{(1)}$ magnetic one-form symmetry likewise emerges.

 With this matter content, the massless theory in the infrared has exact two-group symmetry, and the unification in Table \ref{tab:toyUVExact} gives the maximal possible flavor symmetry consistent with the two-group emergence theorem reviewed in section \ref{sec:emergence}.

{\setlength{\tabcolsep}{0.6 em}
\renewcommand{\arraystretch}{1.3}
\begin{table}[h!]\centering
\large
\begin{tabular}{|c|c|c|}  \hline
 & $SU(2)_g$ & $SU(N_f)$ \\ \hline
$\chi_i = (\chi^+_i, \chi^-_i) $ & $\Box$ & $\Box$ \\ \hline
\end{tabular}\caption{UV completion of toy model with exact flavor symmetry.}\label{tab:toyUVExact}
\end{table}}

Now instead of considering the massless theory, we turn on gauge-invariant masses in the infrared: 
\begin{equation}
    \mathcal{L} \supset y_{ij} m \chi^{+i}\chi^{-j}~.
\end{equation}
This explicitly violates the separate $SU(N_f)_L\times SU(N_f)_R$ symmetries, and the only remaining global symmetries are separate $U(1)$ symmetries acting `axially' on each $\chi^+_i$ and $\chi^-_i$ as is obvious in the mass basis. While the $SU(N_f)$ symmetries are now only approximate, having been broken explicitly by the masses, we can still perform the same two-group analysis as above. Again we may be led to the familiar UV completion, with the flavor symmetry now also approximate in the UV as from a Yukawa coupling
\begin{equation}
    \mathcal{L} \supset y_{ij} \Phi^{ab} \chi^{i}_a\chi^{j}_b~,
\end{equation}
where $a,b$ are $SU(2)_g$ indices and $i,j$ flavor indices. 

However, now that the higher-group symmetry is only approximate, we must take into account the possibility of additional matter becoming light in the limit $y_{\ j}^{i}\rightarrow 0$ and modifying the two-group structure constants.   

{\setlength{\tabcolsep}{0.6 em}
\renewcommand{\arraystretch}{1.3}
\begin{table}[h!]\centering
\large
\begin{tabular}{|c|c|c|}  \hline
 & $SU(2)_g$ & $SU(N_f)$ \\ \hline
$\rho^i = (\chi^{+i}, \psi^{-i}) $ & $\Box$ & $\Box$ \\ 
$\eta_i = (\psi^+_i, \chi^-_i) $ & $\Box$ & $\overline \Box$ \\ \hline
\end{tabular}\caption{UV completion of toy model with approximate infrared flavor symmetry.}\label{tab:toyUVApprox}
\end{table}}

Indeed, we may instead have a UV theory containing additional fermions where $\chi^+_i$ and $\chi^-_i$ now appear in different $SU(2)_g$ multiplets in the UV, as in Table \ref{tab:toyUVApprox}. We can then write down a Yukawa interaction in the UV:
\begin{equation}
    \mathcal{L}\supset y^j_{\ i} \Phi^{ab} \rho^i_{\ a} \eta_{jb}~.
\end{equation}
This Yukawa coupling $y^j_{\ i}$ is a spurion for $SU(N_f)$ symmetry-breaking. After higgsing $SU(2)_g \rightarrow U(1)_g$, we find mass terms and Yukawa interactions:
\begin{equation}
    \mathcal{L}\supset y^j_{\ i} \Lambda \left(\chi^{+i} \chi^{-}_j + \psi^{+i} \psi^{-}_j \right)+y^j_{\ i} \phi \left(\chi^{+i} \chi^{-}_j + \psi^{+i} \psi^{-}_j \right)~,
\end{equation}
which describe the IR theory of Table \ref{tab:toyIRApprox} (together with a neutral scalar $\phi$). Now with respect to our original theory, this UV completion preserves, at least approximately, the \textit{axial} subgroup of the $SU(N_f)_L\times SU(N_f)_R$ symmetry of the $\chi^+, \chi^-$ fields, in which $\chi^+$ transforms as a fundamental and $\chi^-$ as an antifundmental. This is in contrast to the prior UV completion, which by placing the two into the same UV multiplet preserved the exact \textit{vector} subgroup of the massless infrared theory.

{\setlength{\tabcolsep}{0.6 em}
\renewcommand{\arraystretch}{1.3}
\begin{table}[h!]\centering
\large
\begin{tabular}{|c|c|c|c|}  \hline
 & $U(1)_g$ & $SU(N_f)_L$ & $SU(N_f)_R$  \\ \hline
$\chi^+$ & $+1$ & $\Box$ & -- \\
$\chi^-$ & $-1$ & -- & $\overline \Box$ \\
$\psi^+$ & $+1$ & $\Box$ & -- \\
$\psi^-$ & $-1$ & -- & $\overline \Box$ \\ \hline
\end{tabular}\caption{Infrared vector-like abelian gauge theory with new matter.}\label{tab:toyIRApprox}
\end{table}}

We see that we have been able to achieve this by adding infrared matter which is vector-like under the gauge symmetry and chiral under the separate $SU(N_f)_L,  SU(N_f)_R$ symmetries. This also clarifies the sense in which the infrared Yukawas which explicitly violate the would-be flavor symmetries are spurions for the breaking of the two-group structure. Including additional matter which is chiral under the flavor symmetries does not happen at arbitrarily large scales, despite the fact that these new fields have vector-like gauge charges. 

The analysis of Section \ref{sec:emergence} of approximate two-group symmetry tells us that this is in fact a universal feature. In the limit $y_{\ j}^{i}\rightarrow 0$ the full $SU(N_{f})_{L}\times SU(N_{f})_{R}$ flavor symmetry is restored 
 and the two-group analysis robustly shows that  $\chi^{+}$, and $\chi^{-}$ cannot be placed into the same UV multiplet  consistently with this flavor symmetry. Hence the masses of the partner species $\psi^{\pm}$ must also vanish in this limit, so that our analysis of the infrared flavor symmetries and two-group structure constants are modified by them.

In this example we see the proportionality of the vector-like partners' masses to the original species quite directly, because our infrared spectrum is also vector-like. Then the new species have just the same masses as the ones we started out knowing about. 
However, this is the sense in which this toy model does not contain the richness of the Standard Model, and we will see a more interesting version of this below. 

As particle physicists we know now the entire spectrum of species which are \textit{chiral} under the Standard Model gauge group, and so get mass only after electroweak symmetry-breaking. 
The same phenomenon of new vector-like matter which contributes to the chiral flavor symmetries may occur in models of Standard Model unification, but in a far more interesting fashion.
Vector-like matter under the Standard Model gauge group will generically get masses at the GUT scale. Yet because the Standard Model Yukawas are spurions of two-group symmetry breaking we will see by the same argument that such species which participate in Standard Model flavor symmetries must instead appear at the (potentially much lower) scale $y^{\alpha} \Lambda$ with $\Lambda$ the GUT scale and $y$ a Standard Model Yukawa coupling.  We exhibit an explicit example of this in section \ref{sec:trin} below.

\section{The Standard Model}

\subsection{Flavor Symmetry} \label{sec:zeroformflav}

{\setlength{\tabcolsep}{0.6 em}
\renewcommand{\arraystretch}{1.3}
\begin{table}[h!]\centering
\large
\begin{tabular}{|c|c|c|c|c|c|c|c|}  \hline
 & $  Q_i  $ & $ \overline u_i $ & $ \overline d_i $ & $  L_i  $ & $ \overline e_i $ & $  N_i  $ & $ H $ \\ \hline
$SU(3)_C$ & $\mathbf{3}$ & $\overline{\mathbf{3}}$ & $\overline{\mathbf{3}}$ & -- & -- & -- & --\\ \hline
$SU(2)_L$ & $\mathbf{2}$ & -- & -- & $\mathbf{2}$ & -- & -- & $\mathbf{2}$\\ \hline
$U(1)_{Y}$ & $+1$ & $-4$ & $+2$ & $-3$ & $+6$ & -- & $-3$\\ \hline
$U(1)_{B}$ & $+1$ & $-1$ & $-1$ & -- & -- & -- & --\\ \hline
$U(1)_{L}$ & -- & -- & -- & $+1$ & $-1$ & $-1$ & --\\ \hline 
\end{tabular}\caption{Representations of the Standard Model Weyl fermions under the classical gauge and global symmetries. We normalize each $U(1)$ so the least-charged particle has unit charge. We list also the charges of the right-handed neutrino $N$ and the Higgs boson $H$.  }\label{tab:charges}
\end{table}}

By `flavor symmetry of the Standard Model' we will mean a global symmetry of the matter content---not necessarily one which relates different generations of fermions.
That is, a single free, massless Weyl fermion $\psi$ enjoys a classical $U(1)$ `flavor' symmetry, $\psi \rightarrow e^{i \alpha} \psi$. 
A single multiplet of Weyl fermions in a gauge theory without additional interactions continues to enjoy such a flavor symmetry, and we recall the Standard Model gauge structure in Table \ref{tab:charges}, where we label the identical copies of each species $i=1,2,3$, with $N_g=3$ the number of generations in the Standard Model. 
Having this multiplicity of each species upgrades the $U(1)$ symmetry of a single fermion into a $U(3)$ symmetry acting on the fermions of a given species which allows arbitrary unitary rotations among them. The gauge theory does not distinugish between fermions with the same gauge charges.

At this level the Standard Model fields enjoy a large classical $U(3)^5$ flavor symmetry, but to get to the Standard Model we must turn on the Yukawa interactions,
\begin{equation}
    \mathcal{L} \supset y_{ij}^u \tilde{H} Q_i \bar u_j + y_{ij}^d H Q_i \bar d_j + y^e_{ij} H L_i \bar e_j~,
\end{equation}
whose couplings carry the spurion assignments of Table \ref{tab:spurion}. These Yukawas break the classical flavor symmetry explicitly to 
\begin{equation}
     U(1)_{L_e}\times U(1)_{L_\mu} \times U(1)_{L_\tau}\times \frac{U(1)_{B}}{\mathbb{Z}_{3}}~.
\end{equation}
Here $L_i$ are the individual lepton flavor numbers which exist in the Standard Model with massless neutrinos, and the $\mathbb{Z}_3$ quotient accounts for the overlap of baryon number with the center of $SU(3)_C$.\footnote{If we were to include right-handed neutrinos to provide Dirac masses, this would generically change the breaking pattern to $U(3)^6 \rightarrow U(1)_L \times \frac{U(1)_{B}}{\mathbb{Z}_{3}}$, with the similar structure between the quarks and leptons now manifest.} 

{\setlength{\tabcolsep}{0.6 em}
\renewcommand{\arraystretch}{1.3}
\begin{table}[h!]\centering
\large
\begin{tabular}{|c|c|c|c|}  \hline
 & $y^u$ & $y^d$ & $y^e$  \\ \hline
$U(3)_Q$ & $\Box$ & $\Box$ & -- \\
$U(3)_u$ & $ \overline \Box$ & -- & -- \\
$U(3)_d$ & -- & $ \overline \Box$ & -- \\
$U(3)_L$ & -- & -- & $\Box$ \\
$U(3)_e$ & -- & -- & $ \overline \Box$ \\ \hline
\end{tabular}\caption{Traditional zero-form flavor symmetry spurion analysis of the Standard Model Yukawas.}\label{tab:spurion}
\end{table}}

The nonzero Yukawa couplings of the Standard Model break the separate flavor symmetries of each species down to down to a small diagonal subgroup. Nonetheless, as in the toy model above, understanding the higher group structure of the Standard Model flavor symmetries tightly constrains the patterns of unification. And we will see that, as in the toy model above, the presence of symmetry-violation will lead to interesting additional possibilities for unification.

\subsection{Flavor-Hypercharge Two-Group}

As discussed above, triangle diagrams with gauge-global-global legs carry powerful information, despite that they do not lead to violation of the global symmetries around the vacuum. 
In the Standard Model itself there are such nonvanishing triangle coefficients between the flavor symmetries and $U(1)_Y$ hypercharge.
We focus on the $SU(3)_f$ subgroups of the flavor symmetries and list these anomaly coefficients $\kappa_0^{(i)}$ in Table \ref{tab:anomHigher} computed elementarily through \eqref{weylsumform}.

{\setlength{\tabcolsep}{0.6 em}
\renewcommand{\arraystretch}{1.3}
\begin{table}[h!]\centering
\large
\begin{tabular}{|c|c|}  \hline
$\text{Flavor}^{2}$ & $U(1)_Y$ \\ \hline
$SU(3)_Q^2$ & $+ 1 \cdot 2 \cdot N_c$ \\ 
$SU(3)_u^2$ & $-4 \cdot N_c$ \\ 
$SU(3)_d^2$ & $+ 2 \cdot N_c$ \\ 
$SU(3)_L^2$ & $-3 \cdot 2$ \\ 
$SU(3)_e^2$ & $+6$ \\ \hline
\end{tabular}\caption{gauge-global-global triangle coefficients of Standard Model with Yukawas turned off. We note the $U(1)$ part of each flavor symmetry shares the same such coefficients with hypercharge with an additional factor of $N_g$ coefficient, as they have arisen from $U(3) \simeq (SU(3)\times U(1))/\mathbb{Z}_3$.}\label{tab:anomHigher}
\end{table}}

From these coefficients, we see that each individual Standard Model species' flavor symmetry is part of the two-group symmetry.\footnote{This observation was also made in unpublished work of C. C\'{o}rdova, T. Dumitrescu, and K. Intriligator \cite{CDIunpub}.}  
The two-group emergence theorem then tells us that any of these zero-form symmetries whose currents could activate a magnetic one-form symmetry current, as in \eqref{eq:oneOneTwo} must only be a good symmetry below the scale at which the hypercharge magnetic one-form symmetry itself becomes a good symmetry. 

A first statement is then that, contrary to the freedom one often expects in model-building, there is no UV completion of the Standard Model without Yukawa couplings where hypercharge is embedded as, say, $U(1)_Y \subset SU(N)_Y$ and each Standard Model species keeps its own flavor symmetry in the UV theory. 
This simply cannot be accomplished with only the Standard Model spectrum of chiral fermions without turning on Yukawa interactions and allowing additional vector-like matter, as in the toy model above. 
To keep the charged leptons in their own multiplets in such a UV theory would require, for example, vector-like leptons to appear at $\sim y_\ell \Lambda \ll \Lambda$.

Thankfully, such complications need not appear in every unification scheme, as from Table \ref{tab:anomHigher} one observes that the coefficients for Standard Model fields are anything but random, all falling in $\kappa_i = \lbrace -12, -6, 6 \rbrace$. Correspondingly, this means that if we examine diagonal subgroups of multiple species' flavor symmetries, there are subgroups which are not part of the two-group with hypercharge. We count these in Table \ref{tab:non2group} by the number of factors involved. It is these subgroups which may be good symmetries of a UV theory of unification, and in the next section we will exhibit that they are realized in familiar GUTs. While much has been written on the topic of searching for exotic GUTs (e.g. \cite{Okubo:1977sc,Frampton:1979cw,Slansky:1981yr, Eichten:1982pn,Geng:1989tcu,Fishbane:1984zv,Batra:2005rh,Cebola:2014qfa,Yamatsu:2015npn,Feger:2015xqa,Fonseca:2020vke,Feger:2019tvk} as a small sample), our analysis via UV flavor symmetries provides a novel organizing principle.
{\setlength{\tabcolsep}{0.5 em}
\renewcommand{\arraystretch}{1.3}
\begin{table}[h!]\centering
\large
\begin{tabular}{|c|c|c|c|}  \hline
One & \multicolumn{3}{c|}{None} \\ \hline
Two & $\lbrace L, Q\rbrace$ & $\lbrace L, \bar d\rbrace$ & $\lbrace L, \bar e\rbrace$ \\  \hline

Three & $\lbrace \bar u, \bar d, \bar e\rbrace$ & $\lbrace \bar u, \bar e, Q \rbrace$ & $\lbrace \bar u, \bar d, Q\rbrace$   \\  \hline
Four & \multicolumn{3}{c|}{None}  \\  \hline
Five &\multicolumn{3}{c|}{$\lbrace Q, \bar u, \bar d, L, \bar e \rbrace$}\\ \hline

\end{tabular}\caption{Flavor subgroups not participating in the two-group with $U(1)_Y^{(1)}$, indexed by how many fermion species they act on.}\label{tab:non2group}
\end{table}}

Our main focus below will be on classic `vertical' unification schemes where the UV theory continues to have three identical generations, and flavor remains as a global symmetry, by virtue of our focus on the $SU(N_g)_f$ nonabelian flavor subgroups. However, it is possible to extend the analysis to more general cases by considering the gauge-global-global anomalies of $U(1)$ subgroups which are flavor non-universal. Noting that a $U(1)$ acting on a single fermion has the same anomaly coefficient as in Table \ref{tab:anomHigher}, there are then many options for novel unification schemes which one may read off---for example it is allowed to unify the right-handed fermions into the patterns $\lbrace\bar u_1, \bar e_1, \bar e_2 \rbrace$, $\lbrace  \bar u_2, \bar d_1, \bar d_2 \rbrace$ and $\lbrace\bar u_3, \bar d_3, \bar e_3 \rbrace$. But finding a UV theory which intertwines flavor with the Standard Model gauge group in such a way to embed the fermions in these multiplets is a challenge we leave to future study.

Similarly, we need not have imagined restoring the full flavor flavor symmetries with $y^e_{ij}, y^d_{ij}, y^u_{ij} \rightarrow 0$. For example we could study by itself the theory with massless leptons $y^e_{ij} \rightarrow 0$, or keep the third generation massive and consider the flavor symmetries of the first two massless generations.

\section{Unification and the Two-Group} \label{sec:2group}

\subsection{Unification Patterns in Well-Known GUTs}

How can we understand the structure of flavor symmetries in Table \ref{tab:non2group}, and what does it tell us? With only the Standard Model fermions, any subgroup not in this list---any subgroup which is part of the two-group in the IR---must be an emergent accidental symmetry below the scale at which $U(1)_Y$ is unified into a group with no magnetic one-form symmetry. It is only the subgroups in this list that may be good UV symmetries. 
We note again that the gauge-global-global anomaly \eqref{weylsumform} does not distinguish between diagonal subgroups which place different species in fundamentals vs. antifundamentals. That is, two fermions may share the same flavor symmetry in the UV either because they are unified in the same multiplet, or because a Yukawa interaction ties together their possible symmetry transformations.

The top line of Table \ref{tab:non2group} is what we said above: a rotation of one species of fermion independently of any of the others is involved in the two-group with $U(1)_Y$. Any UV completion of hypercharge will invariably break flavor symmetries. The fourth row has the same information and is dual because of the fifth row. The cubic hypercharge anomaly with gravity counts the number of all the degrees of freedom weighted by their hypercharges, and vanishes in the Standard Model:
\begin{equation}
    U(1)_Y \times \text{grav}^2 = \sum_{{\rm species} \ i} g_i Y_i = 0~.
\end{equation}
This is the same counting as for the gauge-global-global coefficient of the fully diagonal flavor subgroup rotating all fermions, so this subgroup does not participate in the two-group with hypercharge. 
Then subgroups which are not part of the two-group partition the fermions---that is, for $\mathfrak{g}$ a subset of Standard Model fermions,
\begin{equation}
    \sum_{i \in \mathfrak{g}} Y_i = 0 \Rightarrow \sum_{i \notin \mathfrak{g}} Y_i = 0~,
\end{equation}
so the $n^{\rm th}$ line and $(5-n)^{\rm th}$ line contain the same information.
In particular, the vanishing gravitational anomaly ensures that we can have a UV theory where simultaneous rotations of all fermions is a UV symmetry because they are all in the same representation---this is what happens in the celebrated $SO(10)$ grand unified theory, \cite{Fritzsch:1974nn}
\begin{equation}
    SO(10): \qquad \psi_{16} = \lbrace Q, \bar u, \bar d, L, \bar e, N \rbrace~,
\end{equation}
where the Standard Model fermions are unified along with the right-handed neutrinos into the 16-dimensional spinor representation.

Of course, it is quite special that the $SO(10)$ representation contains \textit{only} the Standard Model fermions, and that is not generally guaranteed. Indeed, if we instead have a larger UV gauge group such as $E_6$, while this still realizes the option of having all the Standard Model fermions in the same irreducible representation (here the 27), they now come along with additional vector-like fermions. We see also the pattern realized in Georgi-Glashow's \cite{Georgi:1974sy} $SU(5) \subset SO(10)$ as the middle column\footnote{We note that this possibility was not guaranteed by any such cubic gauge anomaly, so its existence is in some sense an accident. We observe also that repeating the exercise with right-handed neutrinos and gauged baryon minus lepton number one finds flavor intertwined in such a way that this $SU(5)$ partitioning is no longer an option, and the only partitions which exist are those guaranteed by the nonabelian gauge groups. Of course, this agrees with the fact that one cannot fit $B-N_cL$ into $SU(5)$.}
\begin{equation}
    {\rm SU(5)}: \qquad \psi_{\bar 5} = \lbrace L, \bar d \rbrace~, \qquad \psi_{10} = \lbrace \bar u, \bar e, Q \rbrace~.
\end{equation}

Turning to the other options of placing the Standard Model fermions into two irreducible representations in the UV, the vanishing of the cubic anomaly $U(1)_Y \times SU(2)_L^2$ also guarantees a partitioning of the degrees of freedom by their $SU(2)_L$ charges. We may unify $\lbrace Q, L\rbrace$ and separately $\lbrace\bar u,\bar d, \bar e, N \rbrace$, as neither of these subgroups participates in the two-group with hypercharge. This is realized in the Pati-Salam model \cite{Pati:1974yy}
\begin{equation}
    G_{\rm PS} \equiv SU(4)_C\times SU(2)_L \times SU(2)_R~,
\end{equation}
where $SU(2)_L$ is untouched and the `left-handed' and `right-handed' fermions are unified separately
\begin{equation}
    \psi_L = (4,2,1) = \lbrace Q, L\rbrace~, \quad \psi_R = (\bar 4, 1, 2) = \lbrace \bar u, \bar d, \bar e, N \rbrace~.
\end{equation}
$G_{\rm PS}$ is also a subgroup of $SO(10)$, places leptons as the `fourth color' of quarks from $SU(4)_C \rightarrow SU(3)_C \times U(1)_{B-N_cL}$, and manifests explicit left-right symmetry.

\subsection{Quark-Lepton Disunification}\label{sec:trin}

By the same logic with which we identified the pattern of Pati-Salam, the vanishing cubic anomaly $U(1)_Y \times SU(3)_C^2$ tells us there may exist a UV flavor symmetry which acts together on the colored Standard Model species $\lbrace Q, \bar u, \bar d \rbrace$ and separately on the uncolored ones $\lbrace L, \bar e, N \rbrace$. 

Such a unification pattern is particularly interesting from the perspective of not having yet observed proton decay---if quarks and leptons are instead unified into the same gauge multiplets in the UV theory, then the heavy vector bosons of the broken directions are necessarily leptoquarks which destabilize the proton in the infrared.
This is the only possible unification pattern which keeps separate the quarks and leptons and so can preserve the proton stability of the Standard Model \cite{Koren:2022bam,Wang:2022eag}, so the stakes are high for properly understanding how this pattern of unification may be realized.

However, this pattern is \textit{not} realized in any theories of unification containing solely the Standard Model species, as checked exhaustively in \cite{Allanach:2021bfe}.
Nonetheless, this UV pattern of flavor symmetries is realized nontrivially in trinification \cite{Babu:1985gi}, which is a product group unification with
\begin{equation}
    G_{3^3} \equiv SU(3)_C \times SU(3)_L \times SU(3)_R~,
\end{equation}
and the fermions contained in
\begin{equation} \label{eq:trinReps}
    \psi_Q = (3, \bar 3, 1)~, \quad \psi_{\bar Q} =  (\bar 3, 1, 3)~, \quad  \psi_L = (1,3,\bar 3)~.
\end{equation}
The quarks in this scheme are split between the left-handed ones in $\psi_Q$ and the right-handed ones in $\psi_{\bar Q}$. Yet Table \ref{tab:non2group} does not seem to allow $\lbrace Q \rbrace, \lbrace \bar u, \bar d \rbrace, \lbrace L, \bar e, N \rbrace$. Indeed, the trinification theory with no Yukawa couplings does \textit{not} lead to the infrared Standard Model flavor symmetry, in accordance with the demands of the two-group structure. Yet trinification theories can reproduce the real Standard Model with its broken flavor symmetries (for recent work thereon see e.g. \cite{Willenbrock:2003ca,Sayre:2006ma,Pelaggi:2015kna,Camargo-Molina:2016bwm,Camargo-Molina:2017kxd,Wang:2018yer}) in much the same manner as we saw in the toy model above.

Approaching the theory from the bottom-up, trinification introduces to the Standard Model chiral matter content an additional vector-like down quark $\lbrace d', \bar d'\rbrace$.  Then there are additional flavor subgroups in the Standard Model $+ \ \lbrace d', \bar d'\rbrace$ which do not participate in the two-group with hypercharge. Namely, those flavor transformations under which the `left-handed' $d'$ transform the same way as the $Q$ flavor multiplet and the `right-handed' $\bar d'$ transform as the $\bar u, \bar d$ flavor multiplets. These expanded flavor symmetries may then be good symmetries in the UV, and indeed they match onto the separate $U(1)$ flavor symmetries of $\psi_Q = \lbrace Q, d' \rbrace$ and $\psi_{\bar Q} = \lbrace \bar u, \bar d, \bar d' \rbrace$. But the theory with these exact flavor symmetries does not produce the Standard Model at low energies. 

Trinification is allowed as a viable unification scheme specifically because the Standard Model flavor symmetries are not exact. Since our extra vector-like quark has been split between the two different UV quark multiplets, we must lift the extra matter by introducing UV couplings which explicitly violate the as-yet-independent flavor symmetries of $\psi_Q$ and $\psi_{\bar Q}$ down to the diagonal. We may return to the Standard Model spectrum by adding a coupling
\begin{equation}
    \mathcal{L} \supset y \Phi \psi_Q \psi_{\bar Q} \supset y \Lambda d' \bar d' + y \tilde{H} Q \bar u + y H Q \bar d~,
\end{equation}
with a field $\Phi$ whose vev $\langle \Phi \rangle = \Lambda$ will break the trinification gauge symmetry and lift the vector-like down quark which we know is absent in the infrared Standard Model. This UV coupling clearly brings the theory back into accord with the two-group emergence theorem, as the only flavor symmetries preserved in the UV now act on all $\lbrace Q, \bar u, \bar d \rbrace$ as we knew they must.

So trinification explicitly realizes the fact that the higher symmetry structure may be modified as a result of the zero-form symmetry being approximate. The Standard Model quark Yukawas act as spurions not only for the zero-form flavor symmetry, but also for the two-group structure, and the modification of this structure is necessarily proportional to these Yukawas.

In fact there is a similar story for the lepton flavor symmetry, but here involving the modification of the cubic 't Hooft anomaly $SU(3)^3$ of the rotations of leptons (we include a right-handed neutrino as an infrared lepton for the purposes of this discussion). The Standard Model flavor subgroup in which all of $\lbrace L, \bar e, N \rbrace$ transform as fundamentals has an anomaly of $4$. Since we place the leptons into the $(1,3,\bar 3)$ of $G_{3^3}$, we must introduce additional leptons which are vector-like under the Standard Model gauge group to fill out this representation, $\psi_L=\lbrace L,\bar e, N, L', \bar L', N'\rbrace$. The additional lepton doublet $L'$, its conjugate $\bar L'$, and an extra sterile neutrino $N'$ are chiral with respect to the $SU(3)$ flavor symmetry of $\psi_L^i$ in the UV theory, so change the cubic anomaly from $4$ to $9$. Since the $SU(3)$ flavor symmetry of the infrared is approximate this is allowed, but it implies that their masses must be proportional to the Standard Model Yukawas which break this symmetry. And indeed, in trinification one adds a Yukawa coupling which is bi-linear in $\psi_L$,
\begin{equation}
    \mathcal{L} \supset y' \Phi' \psi_L \psi_L~.
\end{equation}
This lifts the extra vector-like lepton doublet and leads to seesaw masses for the neutrinos (though only at one-loop due to additional subtleties \cite{Babu:1985gi,Sayre:2006ma}). 
Then while trinification does not modify the two-group structure of the lepton flavor symmetries, it still does modify their 't Hooft anomaly structure, leading again to vector-like leptons with masses suppressed by a Yukawa coupling from the unification scale.

\subsection{Unification Without Monopoles}

As a final example let us consider the pattern of `flipped $SU(5)$' models \cite{Barr:1981qv,Barr:1988yj,Derendinger:1983aj}, which have the structure
\begin{equation}
    SU(5)\times U(1)_\chi~,
\end{equation} 
where the UV theory also has a one-form magnetic symmetry which overlaps with hypercharge. 
The emergence theorem then does not apply, and such theories are indeed able to mix up the Standard Model fermions. Here the fermions are placed into 
\begin{equation}
    \lbrace L, \bar u \rbrace \in (\bar 5, -3), \quad \lbrace Q, \bar d, N \rbrace \in (10,1), \quad \lbrace \bar e \rbrace \in (1,5)~,
\end{equation}
and we see the `flipping' of $\bar e \leftrightarrow N, \bar u \leftrightarrow \bar d$ from the $SU(5)$ multiplet structure. 

This is allowed because the hypercharge magnetic one-form symmetry does not itself emerge at the breaking scale, but rather has some overlap with $U(1)_\chi$. And in fact just the right overlap for the gauge-global-global coefficients to properly match, with the coefficients given in Table \ref{tab:anomFlipped}. In this model hypercharge is embedded as $Y/6 = \chi/5 - Z/5$, where $Z$ is the $SU(5)$ generator which commutes with its $SU(3)_C$ and $SU(2)_L$ subgroups, and indeed with this relation the $U(1)_\chi$-flavor-flavor anomalies in the UV match those of $U(1)_Y$ in the Standard Model, as they must. 

{\setlength{\tabcolsep}{0.6 em}
\renewcommand{\arraystretch}{1.3}
\begin{table}[h]\centering
\large
\begin{tabular}{|c|c|}  \hline
Flavor$^2$ & $U(1)_\chi$ \\ \hline
$SU(3)_5^2$ & $-3 \cdot 5$ \\
$SU(3)_{10}^2$ & $+1 \cdot 10$ \\
$SU(3)_1^2$ & $+5 \cdot 1$ \\ \hline
\end{tabular}\caption{gauge-global-global anomalies of flipped $SU(5)$.}\label{tab:anomFlipped}
\end{table}}

If one were drawn to UV theories in which the $L$ and the $\bar u$ appear in the same multiplet, one could attempt to construct such models by adding additional vector-like matter and using the fact that the Standard Model flavor symmetries are approximate. 
For example by introducing some conjugate Weyl fermions $\lbrace \Psi, \bar \Psi \rbrace$ to spread across two representations as $\lbrace L, \bar u, \bar e, \Psi \rbrace$, and $\lbrace Q, \bar d, N, \bar \Psi \rbrace$.
Then so long as $\Psi$ contributes to the two-group structure constants as $SU(3)_\Psi^2 U(1)_Y = +12$ and $\bar \Psi$ the opposite, we may search for a UV theory with these two multiplets.  
Of course, in order to return to the Standard Model in the infrared there must be a UV Yukawa coupling $y$ between the two such that the separate flavor symmetries are explicitly broken in the UV back to the diagonal. 
The UV theory would then realize a flavor symmetry in which all the Standard Model species transform, though in conjugate representations. 
The Standard Model fermions in the infrared have flavor-symmetry-breaking Yukawa interactions with coupling $y$, and the extra fermions get a vector-like mass $y \Lambda$ whose size is also controlled by the flavor-violating spurion.

\let\oldaddcontentsline\addcontentsline
\renewcommand{\addcontentsline}[3]{}
\section*{Acknowledgements}
\let\addcontentsline\oldaddcontentsline

We thank K. Harigaya and L.T. Wang for helpful conversations, and S. Hong and K. Ohmori for related collaborations.  CC is grateful to T. Dumitrescu and K. Intriligator for preliminary collaboration on the Standard Model two-group \cite{CDIunpub}.  
CC is supported by the DOE DE-SC0009924 and the Simons Collaboration on Global Categorical Symmetries.
SK is supported by an Oehme Postdoctoral Fellowship from the Enrico Fermi Institute at the University of Chicago.


\let\oldaddcontentsline\addcontentsline
\renewcommand{\addcontentsline}[3]{}
\bibliography{2group}

\begin{thebibliography}{68}%
\makeatletter
\providecommand \@ifxundefined [1]{%
 \@ifx{#1\undefined}
}%
\providecommand \@ifnum [1]{%
 \ifnum #1\expandafter \@firstoftwo
 \else \expandafter \@secondoftwo
 \fi
}%
\providecommand \@ifx [1]{%
 \ifx #1\expandafter \@firstoftwo
 \else \expandafter \@secondoftwo
 \fi
}%
\providecommand \natexlab [1]{#1}%
\providecommand \enquote  [1]{``#1''}%
\providecommand \bibnamefont  [1]{#1}%
\providecommand \bibfnamefont [1]{#1}%
\providecommand \citenamefont [1]{#1}%
\providecommand \href@noop [0]{\@secondoftwo}%
\providecommand \href [0]{\begingroup \@sanitize@url \@href}%
\providecommand \@href[1]{\@@startlink{#1}\@@href}%
\providecommand \@@href[1]{\endgroup#1\@@endlink}%
\providecommand \@sanitize@url [0]{\catcode `\\12\catcode `\$12\catcode
  `\&12\catcode `\#12\catcode `\^12\catcode `\_12\catcode `\%12\relax}%
\providecommand \@@startlink[1]{}%
\providecommand \@@endlink[0]{}%
\providecommand \url  [0]{\begingroup\@sanitize@url \@url }%
\providecommand \@url [1]{\endgroup\@href {#1}{\urlprefix }}%
\providecommand \urlprefix  [0]{URL }%
\providecommand \Eprint [0]{\href }%
\providecommand \doibase [0]{https://doi.org/}%
\providecommand \selectlanguage [0]{\@gobble}%
\providecommand \bibinfo  [0]{\@secondoftwo}%
\providecommand \bibfield  [0]{\@secondoftwo}%
\providecommand \translation [1]{[#1]}%
\providecommand \BibitemOpen [0]{}%
\providecommand \bibitemStop [0]{}%
\providecommand \bibitemNoStop [0]{.\EOS\space}%
\providecommand \EOS [0]{\spacefactor3000\relax}%
\providecommand \BibitemShut  [1]{\csname bibitem#1\endcsname}%
\let\auto@bib@innerbib\@empty
\bibitem [{\citenamefont {Gaiotto}\ \emph {et~al.}(2015)\citenamefont
  {Gaiotto}, \citenamefont {Kapustin}, \citenamefont {Seiberg},\ and\
  \citenamefont {Willett}}]{Gaiotto:2014kfa}%
  \BibitemOpen
  \bibfield  {author} {\bibinfo {author} {\bibfnamefont {D.}~\bibnamefont
  {Gaiotto}}, \bibinfo {author} {\bibfnamefont {A.}~\bibnamefont {Kapustin}},
  \bibinfo {author} {\bibfnamefont {N.}~\bibnamefont {Seiberg}},\ and\ \bibinfo
  {author} {\bibfnamefont {B.}~\bibnamefont {Willett}},\ }\bibfield  {title}
  {\bibinfo {title} {{Generalized Global Symmetries}},\ }\href
  {https://doi.org/10.1007/JHEP02(2015)172} {\bibfield  {journal} {\bibinfo
  {journal} {JHEP}\ }\textbf {\bibinfo {volume} {02}},\ \bibinfo {pages}
  {172}},\ \Eprint {https://arxiv.org/abs/1412.5148} {arXiv:1412.5148 [hep-th]}
  \BibitemShut {NoStop}%
\bibitem [{\citenamefont {Glashow}\ \emph {et~al.}(1970)\citenamefont
  {Glashow}, \citenamefont {Iliopoulos},\ and\ \citenamefont
  {Maiani}}]{Glashow:1970gm}%
  \BibitemOpen
  \bibfield  {author} {\bibinfo {author} {\bibfnamefont {S.~L.}\ \bibnamefont
  {Glashow}}, \bibinfo {author} {\bibfnamefont {J.}~\bibnamefont
  {Iliopoulos}},\ and\ \bibinfo {author} {\bibfnamefont {L.}~\bibnamefont
  {Maiani}},\ }\bibfield  {title} {\bibinfo {title} {{Weak Interactions with
  Lepton-Hadron Symmetry}},\ }\href {https://doi.org/10.1103/PhysRevD.2.1285}
  {\bibfield  {journal} {\bibinfo  {journal} {Phys. Rev. D}\ }\textbf {\bibinfo
  {volume} {2}},\ \bibinfo {pages} {1285} (\bibinfo {year} {1970})}\BibitemShut
  {NoStop}%
\bibitem [{\citenamefont {D'Ambrosio}\ \emph {et~al.}(2002)\citenamefont
  {D'Ambrosio}, \citenamefont {Giudice}, \citenamefont {Isidori},\ and\
  \citenamefont {Strumia}}]{DAmbrosio:2002vsn}%
  \BibitemOpen
  \bibfield  {author} {\bibinfo {author} {\bibfnamefont {G.}~\bibnamefont
  {D'Ambrosio}}, \bibinfo {author} {\bibfnamefont {G.~F.}\ \bibnamefont
  {Giudice}}, \bibinfo {author} {\bibfnamefont {G.}~\bibnamefont {Isidori}},\
  and\ \bibinfo {author} {\bibfnamefont {A.}~\bibnamefont {Strumia}},\
  }\bibfield  {title} {\bibinfo {title} {{Minimal flavor violation: An
  Effective field theory approach}},\ }\href
  {https://doi.org/10.1016/S0550-3213(02)00836-2} {\bibfield  {journal}
  {\bibinfo  {journal} {Nucl. Phys. B}\ }\textbf {\bibinfo {volume} {645}},\
  \bibinfo {pages} {155} (\bibinfo {year} {2002})},\ \Eprint
  {https://arxiv.org/abs/hep-ph/0207036} {arXiv:hep-ph/0207036} \BibitemShut
  {NoStop}%
\bibitem [{\citenamefont {Kapustin}\ and\ \citenamefont
  {Thorngren}(2013)}]{Kapustin:2013uxa}%
  \BibitemOpen
  \bibfield  {author} {\bibinfo {author} {\bibfnamefont {A.}~\bibnamefont
  {Kapustin}}\ and\ \bibinfo {author} {\bibfnamefont {R.}~\bibnamefont
  {Thorngren}},\ }\bibfield  {title} {\bibinfo {title} {{Higher symmetry and
  gapped phases of gauge theories}},\ }\href@noop {} {\  (\bibinfo {year}
  {2013})},\ \Eprint {https://arxiv.org/abs/1309.4721} {arXiv:1309.4721
  [hep-th]} \BibitemShut {NoStop}%
\bibitem [{\citenamefont {Tachikawa}(2020)}]{Tachikawa:2017gyf}%
  \BibitemOpen
  \bibfield  {author} {\bibinfo {author} {\bibfnamefont {Y.}~\bibnamefont
  {Tachikawa}},\ }\bibfield  {title} {\bibinfo {title} {{On gauging finite
  subgroups}},\ }\href {https://doi.org/10.21468/SciPostPhys.8.1.015}
  {\bibfield  {journal} {\bibinfo  {journal} {SciPost Phys.}\ }\textbf
  {\bibinfo {volume} {8}},\ \bibinfo {pages} {015} (\bibinfo {year} {2020})},\
  \Eprint {https://arxiv.org/abs/1712.09542} {arXiv:1712.09542 [hep-th]}
  \BibitemShut {NoStop}%
\bibitem [{\citenamefont {C\'ordova}\ \emph {et~al.}(2019)\citenamefont
  {C\'ordova}, \citenamefont {Dumitrescu},\ and\ \citenamefont
  {Intriligator}}]{Cordova:2018cvg}%
  \BibitemOpen
  \bibfield  {author} {\bibinfo {author} {\bibfnamefont {C.}~\bibnamefont
  {C\'ordova}}, \bibinfo {author} {\bibfnamefont {T.~T.}\ \bibnamefont
  {Dumitrescu}},\ and\ \bibinfo {author} {\bibfnamefont {K.}~\bibnamefont
  {Intriligator}},\ }\bibfield  {title} {\bibinfo {title} {{Exploring 2-Group
  Global Symmetries}},\ }\href {https://doi.org/10.1007/JHEP02(2019)184}
  {\bibfield  {journal} {\bibinfo  {journal} {JHEP}\ }\textbf {\bibinfo
  {volume} {02}},\ \bibinfo {pages} {184}},\ \Eprint
  {https://arxiv.org/abs/1802.04790} {arXiv:1802.04790 [hep-th]} \BibitemShut
  {NoStop}%
\bibitem [{\citenamefont {Benini}\ \emph {et~al.}(2019)\citenamefont {Benini},
  \citenamefont {C\'{o}rdova},\ and\ \citenamefont {Hsin}}]{Benini:2018reh}%
  \BibitemOpen
  \bibfield  {author} {\bibinfo {author} {\bibfnamefont {F.}~\bibnamefont
  {Benini}}, \bibinfo {author} {\bibfnamefont {C.}~\bibnamefont
  {C\'{o}rdova}},\ and\ \bibinfo {author} {\bibfnamefont {P.-S.}\ \bibnamefont
  {Hsin}},\ }\bibfield  {title} {\bibinfo {title} {{On 2-Group Global
  Symmetries and their Anomalies}},\ }\href
  {https://doi.org/10.1007/JHEP03(2019)118} {\bibfield  {journal} {\bibinfo
  {journal} {JHEP}\ }\textbf {\bibinfo {volume} {03}},\ \bibinfo {pages}
  {118}},\ \Eprint {https://arxiv.org/abs/1803.09336} {arXiv:1803.09336
  [hep-th]} \BibitemShut {NoStop}%
\bibitem [{\citenamefont {Choi}\ \emph
  {et~al.}(2022{\natexlab{a}})\citenamefont {Choi}, \citenamefont {Lam},\ and\
  \citenamefont {Shao}}]{Choi:2022jqy}%
  \BibitemOpen
  \bibfield  {author} {\bibinfo {author} {\bibfnamefont {Y.}~\bibnamefont
  {Choi}}, \bibinfo {author} {\bibfnamefont {H.~T.}\ \bibnamefont {Lam}},\ and\
  \bibinfo {author} {\bibfnamefont {S.-H.}\ \bibnamefont {Shao}},\ }\bibfield
  {title} {\bibinfo {title} {{Noninvertible Global Symmetries in the Standard
  Model}},\ }\href {https://doi.org/10.1103/PhysRevLett.129.161601} {\bibfield
  {journal} {\bibinfo  {journal} {Phys. Rev. Lett.}\ }\textbf {\bibinfo
  {volume} {129}},\ \bibinfo {pages} {161601} (\bibinfo {year}
  {2022}{\natexlab{a}})},\ \Eprint {https://arxiv.org/abs/2205.05086}
  {arXiv:2205.05086 [hep-th]} \BibitemShut {NoStop}%
\bibitem [{\citenamefont {C\'{o}rdova}\ and\ \citenamefont
  {Ohmori}(2022)}]{Cordova:2022ieu}%
  \BibitemOpen
  \bibfield  {author} {\bibinfo {author} {\bibfnamefont {C.}~\bibnamefont
  {C\'{o}rdova}}\ and\ \bibinfo {author} {\bibfnamefont {K.}~\bibnamefont
  {Ohmori}},\ }\bibfield  {title} {\bibinfo {title} {{Non-Invertible Chiral
  Symmetry and Exponential Hierarchies}},\ }\href@noop {} {\  (\bibinfo {year}
  {2022})},\ \Eprint {https://arxiv.org/abs/2205.06243} {arXiv:2205.06243
  [hep-th]} \BibitemShut {NoStop}%
\bibitem [{\citenamefont {Hidaka}\ \emph {et~al.}(2021)\citenamefont {Hidaka},
  \citenamefont {Nitta},\ and\ \citenamefont {Yokokura}}]{Hidaka:2020izy}%
  \BibitemOpen
  \bibfield  {author} {\bibinfo {author} {\bibfnamefont {Y.}~\bibnamefont
  {Hidaka}}, \bibinfo {author} {\bibfnamefont {M.}~\bibnamefont {Nitta}},\ and\
  \bibinfo {author} {\bibfnamefont {R.}~\bibnamefont {Yokokura}},\ }\bibfield
  {title} {\bibinfo {title} {{Global 3-group symmetry and 't Hooft anomalies in
  axion electrodynamics}},\ }\href {https://doi.org/10.1007/JHEP01(2021)173}
  {\bibfield  {journal} {\bibinfo  {journal} {JHEP}\ }\textbf {\bibinfo
  {volume} {01}},\ \bibinfo {pages} {173}},\ \Eprint
  {https://arxiv.org/abs/2009.14368} {arXiv:2009.14368 [hep-th]} \BibitemShut
  {NoStop}%
\bibitem [{\citenamefont {Choi}\ \emph
  {et~al.}(2022{\natexlab{b}})\citenamefont {Choi}, \citenamefont {Lam},\ and\
  \citenamefont {Shao}}]{Choi:2022rfe}%
  \BibitemOpen
  \bibfield  {author} {\bibinfo {author} {\bibfnamefont {Y.}~\bibnamefont
  {Choi}}, \bibinfo {author} {\bibfnamefont {H.~T.}\ \bibnamefont {Lam}},\ and\
  \bibinfo {author} {\bibfnamefont {S.-H.}\ \bibnamefont {Shao}},\ }\bibfield
  {title} {\bibinfo {title} {{Non-invertible Time-reversal Symmetry}},\
  }\href@noop {} {\  (\bibinfo {year} {2022}{\natexlab{b}})},\ \Eprint
  {https://arxiv.org/abs/2208.04331} {arXiv:2208.04331 [hep-th]} \BibitemShut
  {NoStop}%
\bibitem [{\citenamefont {C\'{o}rdova}\ \emph
  {et~al.}(2022{\natexlab{a}})\citenamefont {C\'{o}rdova}, \citenamefont
  {Hong}, \citenamefont {Koren},\ and\ \citenamefont
  {Ohmori}}]{Cordova:2022fhg}%
  \BibitemOpen
  \bibfield  {author} {\bibinfo {author} {\bibfnamefont {C.}~\bibnamefont
  {C\'{o}rdova}}, \bibinfo {author} {\bibfnamefont {S.}~\bibnamefont {Hong}},
  \bibinfo {author} {\bibfnamefont {S.}~\bibnamefont {Koren}},\ and\ \bibinfo
  {author} {\bibfnamefont {K.}~\bibnamefont {Ohmori}},\ }\bibfield  {title}
  {\bibinfo {title} {{Neutrino Masses from Generalized Symmetry Breaking}},\
  }\href@noop {} {\  (\bibinfo {year} {2022}{\natexlab{a}})},\ \Eprint
  {https://arxiv.org/abs/2211.07639} {arXiv:2211.07639 [hep-ph]} \BibitemShut
  {NoStop}%
\bibitem [{\citenamefont {Hsin}(2022)}]{Hsin:2022heo}%
  \BibitemOpen
  \bibfield  {author} {\bibinfo {author} {\bibfnamefont {P.-S.}\ \bibnamefont
  {Hsin}},\ }\bibfield  {title} {\bibinfo {title} {{Non-Invertible Defects in
  Nonlinear Sigma Models and Coupling to Topological Orders}},\ }\href@noop {}
  {\  (\bibinfo {year} {2022})},\ \Eprint {https://arxiv.org/abs/2212.08608}
  {arXiv:2212.08608 [cond-mat.str-el]} \BibitemShut {NoStop}%
\bibitem [{\citenamefont {Brennan}\ and\ \citenamefont
  {C\'{o}rdova}(2022)}]{Brennan:2020ehu}%
  \BibitemOpen
  \bibfield  {author} {\bibinfo {author} {\bibfnamefont {T.~D.}\ \bibnamefont
  {Brennan}}\ and\ \bibinfo {author} {\bibfnamefont {C.}~\bibnamefont
  {C\'{o}rdova}},\ }\bibfield  {title} {\bibinfo {title} {{Axions,
  higher-groups, and emergent symmetry}},\ }\href
  {https://doi.org/10.1007/JHEP02(2022)145} {\bibfield  {journal} {\bibinfo
  {journal} {JHEP}\ }\textbf {\bibinfo {volume} {02}},\ \bibinfo {pages}
  {145}},\ \Eprint {https://arxiv.org/abs/2011.09600} {arXiv:2011.09600
  [hep-th]} \BibitemShut {NoStop}%
\bibitem [{\citenamefont {Choi}\ \emph
  {et~al.}(2022{\natexlab{c}})\citenamefont {Choi}, \citenamefont {Lam},\ and\
  \citenamefont {Shao}}]{Choi:2022fgx}%
  \BibitemOpen
  \bibfield  {author} {\bibinfo {author} {\bibfnamefont {Y.}~\bibnamefont
  {Choi}}, \bibinfo {author} {\bibfnamefont {H.~T.}\ \bibnamefont {Lam}},\ and\
  \bibinfo {author} {\bibfnamefont {S.-H.}\ \bibnamefont {Shao}},\ }\bibfield
  {title} {\bibinfo {title} {{Non-invertible Gauss Law and Axions}},\
  }\href@noop {} {\  (\bibinfo {year} {2022}{\natexlab{c}})},\ \Eprint
  {https://arxiv.org/abs/2212.04499} {arXiv:2212.04499 [hep-th]} \BibitemShut
  {NoStop}%
\bibitem [{\citenamefont {Wang}(2022)}]{Wang:2022fzc}%
  \BibitemOpen
  \bibfield  {author} {\bibinfo {author} {\bibfnamefont {J.}~\bibnamefont
  {Wang}},\ }\bibfield  {title} {\bibinfo {title} {{CT or P problem and
  symmetric gapped fermion solution}},\ }\href
  {https://doi.org/10.1103/PhysRevD.106.125007} {\bibfield  {journal} {\bibinfo
   {journal} {Phys. Rev. D}\ }\textbf {\bibinfo {volume} {106}},\ \bibinfo
  {pages} {125007} (\bibinfo {year} {2022})},\ \Eprint
  {https://arxiv.org/abs/2207.14813} {arXiv:2207.14813 [hep-th]} \BibitemShut
  {NoStop}%
\bibitem [{\citenamefont {Hinterbichler}\ \emph {et~al.}(2022)\citenamefont
  {Hinterbichler}, \citenamefont {Hofman}, \citenamefont {Joyce},\ and\
  \citenamefont {Mathys}}]{Hinterbichler:2022agn}%
  \BibitemOpen
  \bibfield  {author} {\bibinfo {author} {\bibfnamefont {K.}~\bibnamefont
  {Hinterbichler}}, \bibinfo {author} {\bibfnamefont {D.~M.}\ \bibnamefont
  {Hofman}}, \bibinfo {author} {\bibfnamefont {A.}~\bibnamefont {Joyce}},\ and\
  \bibinfo {author} {\bibfnamefont {G.}~\bibnamefont {Mathys}},\ }\bibfield
  {title} {\bibinfo {title} {{Gravity as a gapless phase and biform
  symmetries}},\ }\href@noop {} {\  (\bibinfo {year} {2022})},\ \Eprint
  {https://arxiv.org/abs/2205.12272} {arXiv:2205.12272 [hep-th]} \BibitemShut
  {NoStop}%
\bibitem [{\citenamefont {C\'{o}rdova}\ \emph
  {et~al.}(2022{\natexlab{b}})\citenamefont {C\'{o}rdova}, \citenamefont
  {Ohmori},\ and\ \citenamefont {Rudelius}}]{Cordova:2022rer}%
  \BibitemOpen
  \bibfield  {author} {\bibinfo {author} {\bibfnamefont {C.}~\bibnamefont
  {C\'{o}rdova}}, \bibinfo {author} {\bibfnamefont {K.}~\bibnamefont
  {Ohmori}},\ and\ \bibinfo {author} {\bibfnamefont {T.}~\bibnamefont
  {Rudelius}},\ }\bibfield  {title} {\bibinfo {title} {{Generalized symmetry
  breaking scales and weak gravity conjectures}},\ }\href
  {https://doi.org/10.1007/JHEP11(2022)154} {\bibfield  {journal} {\bibinfo
  {journal} {JHEP}\ }\textbf {\bibinfo {volume} {11}},\ \bibinfo {pages}
  {154}},\ \Eprint {https://arxiv.org/abs/2202.05866} {arXiv:2202.05866
  [hep-th]} \BibitemShut {NoStop}%
\bibitem [{\citenamefont {Anber}\ \emph {et~al.}(2022)\citenamefont {Anber},
  \citenamefont {Hong},\ and\ \citenamefont {Son}}]{Anber:2021iip}%
  \BibitemOpen
  \bibfield  {author} {\bibinfo {author} {\bibfnamefont {M.~M.}\ \bibnamefont
  {Anber}}, \bibinfo {author} {\bibfnamefont {S.}~\bibnamefont {Hong}},\ and\
  \bibinfo {author} {\bibfnamefont {M.}~\bibnamefont {Son}},\ }\bibfield
  {title} {\bibinfo {title} {{New anomalies, TQFTs, and confinement in bosonic
  chiral gauge theories}},\ }\href {https://doi.org/10.1007/JHEP02(2022)062}
  {\bibfield  {journal} {\bibinfo  {journal} {JHEP}\ }\textbf {\bibinfo
  {volume} {02}},\ \bibinfo {pages} {062}},\ \Eprint
  {https://arxiv.org/abs/2109.03245} {arXiv:2109.03245 [hep-th]} \BibitemShut
  {NoStop}%
\bibitem [{\citenamefont {Fan}\ \emph {et~al.}(2021)\citenamefont {Fan},
  \citenamefont {Fraser}, \citenamefont {Reece},\ and\ \citenamefont
  {Stout}}]{Fan:2021ntg}%
  \BibitemOpen
  \bibfield  {author} {\bibinfo {author} {\bibfnamefont {J.}~\bibnamefont
  {Fan}}, \bibinfo {author} {\bibfnamefont {K.}~\bibnamefont {Fraser}},
  \bibinfo {author} {\bibfnamefont {M.}~\bibnamefont {Reece}},\ and\ \bibinfo
  {author} {\bibfnamefont {J.}~\bibnamefont {Stout}},\ }\bibfield  {title}
  {\bibinfo {title} {{Axion Mass from Magnetic Monopole Loops}},\ }\href
  {https://doi.org/10.1103/PhysRevLett.127.131602} {\bibfield  {journal}
  {\bibinfo  {journal} {Phys. Rev. Lett.}\ }\textbf {\bibinfo {volume} {127}},\
  \bibinfo {pages} {131602} (\bibinfo {year} {2021})},\ \Eprint
  {https://arxiv.org/abs/2105.09950} {arXiv:2105.09950 [hep-ph]} \BibitemShut
  {NoStop}%
\bibitem [{\citenamefont {Hong}\ and\ \citenamefont
  {Rigo}(2021)}]{Hong:2020bvq}%
  \BibitemOpen
  \bibfield  {author} {\bibinfo {author} {\bibfnamefont {S.}~\bibnamefont
  {Hong}}\ and\ \bibinfo {author} {\bibfnamefont {G.}~\bibnamefont {Rigo}},\
  }\bibfield  {title} {\bibinfo {title} {{Anomaly Inflow and Holography}},\
  }\href {https://doi.org/10.1007/JHEP05(2021)072} {\bibfield  {journal}
  {\bibinfo  {journal} {JHEP}\ }\textbf {\bibinfo {volume} {05}},\ \bibinfo
  {pages} {072}},\ \Eprint {https://arxiv.org/abs/2012.03964} {arXiv:2012.03964
  [hep-th]} \BibitemShut {NoStop}%
\bibitem [{\citenamefont {Brennan}\ \emph {et~al.}()\citenamefont {Brennan},
  \citenamefont {Hong},\ and\ \citenamefont {Wang}}]{Brennan:2023xyz}%
  \BibitemOpen
  \bibfield  {author} {\bibinfo {author} {\bibfnamefont {D.}~\bibnamefont
  {Brennan}}, \bibinfo {author} {\bibfnamefont {S.}~\bibnamefont {Hong}},\ and\
  \bibinfo {author} {\bibfnamefont {L.-T.}\ \bibnamefont {Wang}},\ }\bibfield
  {title} {\bibinfo {title} {{Coupling a Cosmic String to a TQFT}},\
  }\href@noop {} {\ }\Eprint {https://arxiv.org/abs/To appear} {To appear}
  \BibitemShut {NoStop}%
\bibitem [{\citenamefont {Heidenreich}\ \emph {et~al.}(2021)\citenamefont
  {Heidenreich}, \citenamefont {McNamara}, \citenamefont {Montero},
  \citenamefont {Reece}, \citenamefont {Rudelius},\ and\ \citenamefont
  {Valenzuela}}]{Heidenreich:2020pkc}%
  \BibitemOpen
  \bibfield  {author} {\bibinfo {author} {\bibfnamefont {B.}~\bibnamefont
  {Heidenreich}}, \bibinfo {author} {\bibfnamefont {J.}~\bibnamefont
  {McNamara}}, \bibinfo {author} {\bibfnamefont {M.}~\bibnamefont {Montero}},
  \bibinfo {author} {\bibfnamefont {M.}~\bibnamefont {Reece}}, \bibinfo
  {author} {\bibfnamefont {T.}~\bibnamefont {Rudelius}},\ and\ \bibinfo
  {author} {\bibfnamefont {I.}~\bibnamefont {Valenzuela}},\ }\bibfield  {title}
  {\bibinfo {title} {{Chern-Weil global symmetries and how quantum gravity
  avoids them}},\ }\href {https://doi.org/10.1007/JHEP11(2021)053} {\bibfield
  {journal} {\bibinfo  {journal} {JHEP}\ }\textbf {\bibinfo {volume} {11}},\
  \bibinfo {pages} {053}},\ \Eprint {https://arxiv.org/abs/2012.00009}
  {arXiv:2012.00009 [hep-th]} \BibitemShut {NoStop}%
\bibitem [{\citenamefont {McNamara}\ and\ \citenamefont
  {Reece}(2022)}]{McNamara:2022lrw}%
  \BibitemOpen
  \bibfield  {author} {\bibinfo {author} {\bibfnamefont {J.}~\bibnamefont
  {McNamara}}\ and\ \bibinfo {author} {\bibfnamefont {M.}~\bibnamefont
  {Reece}},\ }\bibfield  {title} {\bibinfo {title} {{Reflections on Parity
  Breaking}},\ }\href@noop {} {\  (\bibinfo {year} {2022})},\ \Eprint
  {https://arxiv.org/abs/2212.00039} {arXiv:2212.00039 [hep-th]} \BibitemShut
  {NoStop}%
\bibitem [{\citenamefont {Wang}\ and\ \citenamefont
  {Wen}(2020)}]{Wang:2018jkc}%
  \BibitemOpen
  \bibfield  {author} {\bibinfo {author} {\bibfnamefont {J.}~\bibnamefont
  {Wang}}\ and\ \bibinfo {author} {\bibfnamefont {X.-G.}\ \bibnamefont {Wen}},\
  }\bibfield  {title} {\bibinfo {title} {{Nonperturbative definition of the
  standard models}},\ }\href {https://doi.org/10.1103/PhysRevResearch.2.023356}
  {\bibfield  {journal} {\bibinfo  {journal} {Phys. Rev. Res.}\ }\textbf
  {\bibinfo {volume} {2}},\ \bibinfo {pages} {023356} (\bibinfo {year}
  {2020})},\ \Eprint {https://arxiv.org/abs/1809.11171} {arXiv:1809.11171
  [hep-th]} \BibitemShut {NoStop}%
\bibitem [{\citenamefont {Wan}\ and\ \citenamefont {Wang}(2020)}]{Wan:2019gqr}%
  \BibitemOpen
  \bibfield  {author} {\bibinfo {author} {\bibfnamefont {Z.}~\bibnamefont
  {Wan}}\ and\ \bibinfo {author} {\bibfnamefont {J.}~\bibnamefont {Wang}},\
  }\bibfield  {title} {\bibinfo {title} {{Beyond Standard Models and Grand
  Unifications: Anomalies, Topological Terms, and Dynamical Constraints via
  Cobordisms}},\ }\href {https://doi.org/10.1007/JHEP07(2020)062} {\bibfield
  {journal} {\bibinfo  {journal} {JHEP}\ }\textbf {\bibinfo {volume} {07}},\
  \bibinfo {pages} {062}},\ \Eprint {https://arxiv.org/abs/1910.14668}
  {arXiv:1910.14668 [hep-th]} \BibitemShut {NoStop}%
\bibitem [{\citenamefont {Davighi}\ \emph {et~al.}(2020)\citenamefont
  {Davighi}, \citenamefont {Gripaios},\ and\ \citenamefont
  {Lohitsiri}}]{Davighi:2019rcd}%
  \BibitemOpen
  \bibfield  {author} {\bibinfo {author} {\bibfnamefont {J.}~\bibnamefont
  {Davighi}}, \bibinfo {author} {\bibfnamefont {B.}~\bibnamefont {Gripaios}},\
  and\ \bibinfo {author} {\bibfnamefont {N.}~\bibnamefont {Lohitsiri}},\
  }\bibfield  {title} {\bibinfo {title} {{Global anomalies in the Standard
  Model(s) and Beyond}},\ }\href {https://doi.org/10.1007/JHEP07(2020)232}
  {\bibfield  {journal} {\bibinfo  {journal} {JHEP}\ }\textbf {\bibinfo
  {volume} {07}},\ \bibinfo {pages} {232}},\ \Eprint
  {https://arxiv.org/abs/1910.11277} {arXiv:1910.11277 [hep-th]} \BibitemShut
  {NoStop}%
\bibitem [{\citenamefont {Wang}(2021)}]{Wang:2020mra}%
  \BibitemOpen
  \bibfield  {author} {\bibinfo {author} {\bibfnamefont {J.}~\bibnamefont
  {Wang}},\ }\bibfield  {title} {\bibinfo {title} {{Ultra Unification}},\
  }\href {https://doi.org/10.1103/PhysRevD.103.105024} {\bibfield  {journal}
  {\bibinfo  {journal} {Phys. Rev. D}\ }\textbf {\bibinfo {volume} {103}},\
  \bibinfo {pages} {105024} (\bibinfo {year} {2021})},\ \Eprint
  {https://arxiv.org/abs/2012.15860} {arXiv:2012.15860 [hep-th]} \BibitemShut
  {NoStop}%
\bibitem [{\citenamefont {Wang}(2020)}]{Wang:2020xyo}%
  \BibitemOpen
  \bibfield  {author} {\bibinfo {author} {\bibfnamefont {J.}~\bibnamefont
  {Wang}},\ }\bibfield  {title} {\bibinfo {title} {{Anomaly and Cobordism
  Constraints Beyond the Standard Model: Topological Force}},\ }\href@noop {}
  {\  (\bibinfo {year} {2020})},\ \Eprint {https://arxiv.org/abs/2006.16996}
  {arXiv:2006.16996 [hep-th]} \BibitemShut {NoStop}%
\bibitem [{\citenamefont {Anber}\ and\ \citenamefont
  {Poppitz}(2021)}]{Anber:2021upc}%
  \BibitemOpen
  \bibfield  {author} {\bibinfo {author} {\bibfnamefont {M.~M.}\ \bibnamefont
  {Anber}}\ and\ \bibinfo {author} {\bibfnamefont {E.}~\bibnamefont
  {Poppitz}},\ }\bibfield  {title} {\bibinfo {title} {{Nonperturbative effects
  in the Standard Model with gauged 1-form symmetry}},\ }\href
  {https://doi.org/10.1007/JHEP12(2021)055} {\bibfield  {journal} {\bibinfo
  {journal} {JHEP}\ }\textbf {\bibinfo {volume} {12}},\ \bibinfo {pages}
  {055}},\ \Eprint {https://arxiv.org/abs/2110.02981} {arXiv:2110.02981
  [hep-th]} \BibitemShut {NoStop}%
\bibitem [{\citenamefont {Wang}\ \emph
  {et~al.}(2022{\natexlab{a}})\citenamefont {Wang}, \citenamefont {Wan},\ and\
  \citenamefont {You}}]{Wang:2021ayd}%
  \BibitemOpen
  \bibfield  {author} {\bibinfo {author} {\bibfnamefont {J.}~\bibnamefont
  {Wang}}, \bibinfo {author} {\bibfnamefont {Z.}~\bibnamefont {Wan}},\ and\
  \bibinfo {author} {\bibfnamefont {Y.-Z.}\ \bibnamefont {You}},\ }\bibfield
  {title} {\bibinfo {title} {{Cobordism and deformation class of the standard
  model}},\ }\href {https://doi.org/10.1103/PhysRevD.106.L041701} {\bibfield
  {journal} {\bibinfo  {journal} {Phys. Rev. D}\ }\textbf {\bibinfo {volume}
  {106}},\ \bibinfo {pages} {L041701} (\bibinfo {year} {2022}{\natexlab{a}})},\
  \Eprint {https://arxiv.org/abs/2112.14765} {arXiv:2112.14765 [hep-th]}
  \BibitemShut {NoStop}%
\bibitem [{\citenamefont {Wang}\ and\ \citenamefont
  {You}(2021)}]{Wang:2021vki}%
  \BibitemOpen
  \bibfield  {author} {\bibinfo {author} {\bibfnamefont {J.}~\bibnamefont
  {Wang}}\ and\ \bibinfo {author} {\bibfnamefont {Y.-Z.}\ \bibnamefont {You}},\
  }\bibfield  {title} {\bibinfo {title} {{Gauge Enhanced Quantum Criticality
  Between Grand Unifications: Categorical Higher Symmetry Retraction}},\
  }\href@noop {} {\  (\bibinfo {year} {2021})},\ \Eprint
  {https://arxiv.org/abs/2111.10369} {arXiv:2111.10369 [hep-th]} \BibitemShut
  {NoStop}%
\bibitem [{\citenamefont {Wang}\ and\ \citenamefont
  {You}(2022)}]{Wang:2021hob}%
  \BibitemOpen
  \bibfield  {author} {\bibinfo {author} {\bibfnamefont {J.}~\bibnamefont
  {Wang}}\ and\ \bibinfo {author} {\bibfnamefont {Y.-Z.}\ \bibnamefont {You}},\
  }\bibfield  {title} {\bibinfo {title} {{Gauge enhanced quantum criticality
  beyond the standard model}},\ }\href
  {https://doi.org/10.1103/PhysRevD.106.025013} {\bibfield  {journal} {\bibinfo
   {journal} {Phys. Rev. D}\ }\textbf {\bibinfo {volume} {106}},\ \bibinfo
  {pages} {025013} (\bibinfo {year} {2022})},\ \Eprint
  {https://arxiv.org/abs/2106.16248} {arXiv:2106.16248 [hep-th]} \BibitemShut
  {NoStop}%
\bibitem [{\citenamefont {Davighi}\ \emph {et~al.}(2022)\citenamefont
  {Davighi}, \citenamefont {Gripaios},\ and\ \citenamefont
  {Lohitsiri}}]{Davighi:2022icj}%
  \BibitemOpen
  \bibfield  {author} {\bibinfo {author} {\bibfnamefont {J.}~\bibnamefont
  {Davighi}}, \bibinfo {author} {\bibfnamefont {B.}~\bibnamefont {Gripaios}},\
  and\ \bibinfo {author} {\bibfnamefont {N.}~\bibnamefont {Lohitsiri}},\
  }\bibfield  {title} {\bibinfo {title} {{Anomalies of non-Abelian finite
  groups via cobordism}},\ }\href {https://doi.org/10.1007/JHEP09(2022)147}
  {\bibfield  {journal} {\bibinfo  {journal} {JHEP}\ }\textbf {\bibinfo
  {volume} {09}},\ \bibinfo {pages} {147}},\ \Eprint
  {https://arxiv.org/abs/2207.10700} {arXiv:2207.10700 [hep-th]} \BibitemShut
  {NoStop}%
\bibitem [{\citenamefont {Davighi}\ and\ \citenamefont
  {Lohitsiri}(2021)}]{Davighi:2020uab}%
  \BibitemOpen
  \bibfield  {author} {\bibinfo {author} {\bibfnamefont {J.}~\bibnamefont
  {Davighi}}\ and\ \bibinfo {author} {\bibfnamefont {N.}~\bibnamefont
  {Lohitsiri}},\ }\bibfield  {title} {\bibinfo {title} {{The algebra of anomaly
  interplay}},\ }\href {https://doi.org/10.21468/SciPostPhys.10.3.074}
  {\bibfield  {journal} {\bibinfo  {journal} {SciPost Phys.}\ }\textbf
  {\bibinfo {volume} {10}},\ \bibinfo {pages} {074} (\bibinfo {year} {2021})},\
  \Eprint {https://arxiv.org/abs/2011.10102} {arXiv:2011.10102 [hep-th]}
  \BibitemShut {NoStop}%
\bibitem [{\citenamefont {Choi}\ \emph
  {et~al.}(2022{\natexlab{d}})\citenamefont {Choi}, \citenamefont
  {C\'{o}rdova}, \citenamefont {Hsin}, \citenamefont {Lam},\ and\ \citenamefont
  {Shao}}]{Choi:2021kmx}%
  \BibitemOpen
  \bibfield  {author} {\bibinfo {author} {\bibfnamefont {Y.}~\bibnamefont
  {Choi}}, \bibinfo {author} {\bibfnamefont {C.}~\bibnamefont {C\'{o}rdova}},
  \bibinfo {author} {\bibfnamefont {P.-S.}\ \bibnamefont {Hsin}}, \bibinfo
  {author} {\bibfnamefont {H.~T.}\ \bibnamefont {Lam}},\ and\ \bibinfo {author}
  {\bibfnamefont {S.-H.}\ \bibnamefont {Shao}},\ }\bibfield  {title} {\bibinfo
  {title} {{Noninvertible duality defects in 3+1 dimensions}},\ }\href
  {https://doi.org/10.1103/PhysRevD.105.125016} {\bibfield  {journal} {\bibinfo
   {journal} {Phys. Rev. D}\ }\textbf {\bibinfo {volume} {105}},\ \bibinfo
  {pages} {125016} (\bibinfo {year} {2022}{\natexlab{d}})},\ \Eprint
  {https://arxiv.org/abs/2111.01139} {arXiv:2111.01139 [hep-th]} \BibitemShut
  {NoStop}%
\bibitem [{\citenamefont {Kaidi}\ \emph {et~al.}(2022)\citenamefont {Kaidi},
  \citenamefont {Ohmori},\ and\ \citenamefont {Zheng}}]{Kaidi:2021xfk}%
  \BibitemOpen
  \bibfield  {author} {\bibinfo {author} {\bibfnamefont {J.}~\bibnamefont
  {Kaidi}}, \bibinfo {author} {\bibfnamefont {K.}~\bibnamefont {Ohmori}},\ and\
  \bibinfo {author} {\bibfnamefont {Y.}~\bibnamefont {Zheng}},\ }\bibfield
  {title} {\bibinfo {title} {{Kramers-Wannier-like Duality Defects in (3+1)D
  Gauge Theories}},\ }\href {https://doi.org/10.1103/PhysRevLett.128.111601}
  {\bibfield  {journal} {\bibinfo  {journal} {Phys. Rev. Lett.}\ }\textbf
  {\bibinfo {volume} {128}},\ \bibinfo {pages} {111601} (\bibinfo {year}
  {2022})},\ \Eprint {https://arxiv.org/abs/2111.01141} {arXiv:2111.01141
  [hep-th]} \BibitemShut {NoStop}%
\bibitem [{\citenamefont {Froggatt}\ and\ \citenamefont
  {Nielsen}(1979)}]{Froggatt:1978nt}%
  \BibitemOpen
  \bibfield  {author} {\bibinfo {author} {\bibfnamefont {C.~D.}\ \bibnamefont
  {Froggatt}}\ and\ \bibinfo {author} {\bibfnamefont {H.~B.}\ \bibnamefont
  {Nielsen}},\ }\bibfield  {title} {\bibinfo {title} {{Hierarchy of Quark
  Masses, Cabibbo Angles and CP Violation}},\ }\href
  {https://doi.org/10.1016/0550-3213(79)90316-X} {\bibfield  {journal}
  {\bibinfo  {journal} {Nucl. Phys. B}\ }\textbf {\bibinfo {volume} {147}},\
  \bibinfo {pages} {277} (\bibinfo {year} {1979})}\BibitemShut {NoStop}%
\bibitem [{\citenamefont {Witten}(1982)}]{Witten:1982fp}%
  \BibitemOpen
  \bibfield  {author} {\bibinfo {author} {\bibfnamefont {E.}~\bibnamefont
  {Witten}},\ }\bibfield  {title} {\bibinfo {title} {{An SU(2) Anomaly}},\
  }\href {https://doi.org/10.1016/0370-2693(82)90728-6} {\bibfield  {journal}
  {\bibinfo  {journal} {Phys. Lett. B}\ }\textbf {\bibinfo {volume} {117}},\
  \bibinfo {pages} {324} (\bibinfo {year} {1982})}\BibitemShut {NoStop}%
\bibitem [{\citenamefont {C\'{o}rdova}\ \emph {et~al.}(2018)\citenamefont
  {C\'{o}rdova}, \citenamefont {Dumitrescu},\ and\ \citenamefont
  {Intriligator}}]{CDIunpub}%
  \BibitemOpen
  \bibfield  {author} {\bibinfo {author} {\bibfnamefont {C.}~\bibnamefont
  {C\'{o}rdova}}, \bibinfo {author} {\bibfnamefont {T.}~\bibnamefont
  {Dumitrescu}},\ and\ \bibinfo {author} {\bibfnamefont {K.}~\bibnamefont
  {Intriligator}},\ }\bibfield  {title} {\bibinfo {title} {{Unpublished}},\
  }\href@noop {} {\  (\bibinfo {year} {2018})}\BibitemShut {NoStop}%
\bibitem [{\citenamefont {Okubo}(1977)}]{Okubo:1977sc}%
  \BibitemOpen
  \bibfield  {author} {\bibinfo {author} {\bibfnamefont {S.}~\bibnamefont
  {Okubo}},\ }\bibfield  {title} {\bibinfo {title} {{Gauge Groups Without
  Triangular Anomaly}},\ }\href {https://doi.org/10.1103/PhysRevD.16.3528}
  {\bibfield  {journal} {\bibinfo  {journal} {Phys. Rev. D}\ }\textbf {\bibinfo
  {volume} {16}},\ \bibinfo {pages} {3528} (\bibinfo {year}
  {1977})}\BibitemShut {NoStop}%
\bibitem [{\citenamefont {Frampton}(1979)}]{Frampton:1979cw}%
  \BibitemOpen
  \bibfield  {author} {\bibinfo {author} {\bibfnamefont {P.~H.}\ \bibnamefont
  {Frampton}},\ }\bibfield  {title} {\bibinfo {title} {{SU($N$) Grand
  Unification With Several Quark - Lepton Generations}},\ }\href
  {https://doi.org/10.1016/0370-2693(79)90472-6} {\bibfield  {journal}
  {\bibinfo  {journal} {Phys. Lett. B}\ }\textbf {\bibinfo {volume} {88}},\
  \bibinfo {pages} {299} (\bibinfo {year} {1979})}\BibitemShut {NoStop}%
\bibitem [{\citenamefont {Slansky}(1981)}]{Slansky:1981yr}%
  \BibitemOpen
  \bibfield  {author} {\bibinfo {author} {\bibfnamefont {R.}~\bibnamefont
  {Slansky}},\ }\bibfield  {title} {\bibinfo {title} {{Group Theory for Unified
  Model Building}},\ }\href {https://doi.org/10.1016/0370-1573(81)90092-2}
  {\bibfield  {journal} {\bibinfo  {journal} {Phys. Rept.}\ }\textbf {\bibinfo
  {volume} {79}},\ \bibinfo {pages} {1} (\bibinfo {year} {1981})}\BibitemShut
  {NoStop}%
\bibitem [{\citenamefont {Eichten}\ \emph {et~al.}(1982)\citenamefont
  {Eichten}, \citenamefont {Kang},\ and\ \citenamefont {Koh}}]{Eichten:1982pn}%
  \BibitemOpen
  \bibfield  {author} {\bibinfo {author} {\bibfnamefont {E.}~\bibnamefont
  {Eichten}}, \bibinfo {author} {\bibfnamefont {K.}~\bibnamefont {Kang}},\ and\
  \bibinfo {author} {\bibfnamefont {I.-G.}\ \bibnamefont {Koh}},\ }\bibfield
  {title} {\bibinfo {title} {{Anomaly Free Complex Representations in SU(N)}},\
  }\href {https://doi.org/10.1063/1.525299} {\bibfield  {journal} {\bibinfo
  {journal} {J. Math. Phys.}\ }\textbf {\bibinfo {volume} {23}},\ \bibinfo
  {pages} {2529} (\bibinfo {year} {1982})}\BibitemShut {NoStop}%
\bibitem [{\citenamefont {Geng}\ and\ \citenamefont
  {Marshak}(1989)}]{Geng:1989tcu}%
  \BibitemOpen
  \bibfield  {author} {\bibinfo {author} {\bibfnamefont {C.~Q.}\ \bibnamefont
  {Geng}}\ and\ \bibinfo {author} {\bibfnamefont {R.~E.}\ \bibnamefont
  {Marshak}},\ }\bibfield  {title} {\bibinfo {title} {{Uniqueness of Quark and
  Lepton Representations in the Standard Model From the Anomalies Viewpoint}},\
  }\href {https://doi.org/10.1103/PhysRevD.39.693} {\bibfield  {journal}
  {\bibinfo  {journal} {Phys. Rev. D}\ }\textbf {\bibinfo {volume} {39}},\
  \bibinfo {pages} {693} (\bibinfo {year} {1989})}\BibitemShut {NoStop}%
\bibitem [{\citenamefont {Fishbane}\ \emph {et~al.}(1985)\citenamefont
  {Fishbane}, \citenamefont {Meshkov}, \citenamefont {Norton},\ and\
  \citenamefont {Ramond}}]{Fishbane:1984zv}%
  \BibitemOpen
  \bibfield  {author} {\bibinfo {author} {\bibfnamefont {P.~M.}\ \bibnamefont
  {Fishbane}}, \bibinfo {author} {\bibfnamefont {S.}~\bibnamefont {Meshkov}},
  \bibinfo {author} {\bibfnamefont {R.~E.}\ \bibnamefont {Norton}},\ and\
  \bibinfo {author} {\bibfnamefont {P.}~\bibnamefont {Ramond}},\ }\bibfield
  {title} {\bibinfo {title} {{Chiral Fermions Beyond the Standard Model}},\
  }\href {https://doi.org/10.1103/PhysRevD.31.1119} {\bibfield  {journal}
  {\bibinfo  {journal} {Phys. Rev. D}\ }\textbf {\bibinfo {volume} {31}},\
  \bibinfo {pages} {1119} (\bibinfo {year} {1985})}\BibitemShut {NoStop}%
\bibitem [{\citenamefont {Batra}\ \emph {et~al.}(2006)\citenamefont {Batra},
  \citenamefont {Dobrescu},\ and\ \citenamefont {Spivak}}]{Batra:2005rh}%
  \BibitemOpen
  \bibfield  {author} {\bibinfo {author} {\bibfnamefont {P.}~\bibnamefont
  {Batra}}, \bibinfo {author} {\bibfnamefont {B.~A.}\ \bibnamefont
  {Dobrescu}},\ and\ \bibinfo {author} {\bibfnamefont {D.}~\bibnamefont
  {Spivak}},\ }\bibfield  {title} {\bibinfo {title} {{Anomaly-free sets of
  fermions}},\ }\href {https://doi.org/10.1063/1.2222081} {\bibfield  {journal}
  {\bibinfo  {journal} {J. Math. Phys.}\ }\textbf {\bibinfo {volume} {47}},\
  \bibinfo {pages} {082301} (\bibinfo {year} {2006})},\ \Eprint
  {https://arxiv.org/abs/hep-ph/0510181} {arXiv:hep-ph/0510181} \BibitemShut
  {NoStop}%
\bibitem [{\citenamefont {Cebola}\ \emph {et~al.}(2014)\citenamefont {Cebola},
  \citenamefont {Emmanuel-Costa}, \citenamefont {Gonz\'alez~Felipe},\ and\
  \citenamefont {Sim\~oes}}]{Cebola:2014qfa}%
  \BibitemOpen
  \bibfield  {author} {\bibinfo {author} {\bibfnamefont {L.~M.}\ \bibnamefont
  {Cebola}}, \bibinfo {author} {\bibfnamefont {D.}~\bibnamefont
  {Emmanuel-Costa}}, \bibinfo {author} {\bibfnamefont {R.}~\bibnamefont
  {Gonz\'alez~Felipe}},\ and\ \bibinfo {author} {\bibfnamefont
  {C.}~\bibnamefont {Sim\~oes}},\ }\bibfield  {title} {\bibinfo {title}
  {{Minimal anomaly-free chiral fermion sets and gauge coupling unification}},\
  }\href {https://doi.org/10.1103/PhysRevD.90.125037} {\bibfield  {journal}
  {\bibinfo  {journal} {Phys. Rev. D}\ }\textbf {\bibinfo {volume} {90}},\
  \bibinfo {pages} {125037} (\bibinfo {year} {2014})},\ \Eprint
  {https://arxiv.org/abs/1409.0805} {arXiv:1409.0805 [hep-ph]} \BibitemShut
  {NoStop}%
\bibitem [{\citenamefont {Yamatsu}(2015)}]{Yamatsu:2015npn}%
  \BibitemOpen
  \bibfield  {author} {\bibinfo {author} {\bibfnamefont {N.}~\bibnamefont
  {Yamatsu}},\ }\bibfield  {title} {\bibinfo {title} {{Finite-Dimensional Lie
  Algebras and Their Representations for Unified Model Building}},\ }\href@noop
  {} {\  (\bibinfo {year} {2015})},\ \Eprint {https://arxiv.org/abs/1511.08771}
  {arXiv:1511.08771 [hep-ph]} \BibitemShut {NoStop}%
\bibitem [{\citenamefont {Feger}\ and\ \citenamefont
  {Kephart}(2015)}]{Feger:2015xqa}%
  \BibitemOpen
  \bibfield  {author} {\bibinfo {author} {\bibfnamefont {R.~P.}\ \bibnamefont
  {Feger}}\ and\ \bibinfo {author} {\bibfnamefont {T.~W.}\ \bibnamefont
  {Kephart}},\ }\bibfield  {title} {\bibinfo {title} {{Grand Unification and
  Exotic Fermions}},\ }\href {https://doi.org/10.1103/PhysRevD.92.035005}
  {\bibfield  {journal} {\bibinfo  {journal} {Phys. Rev. D}\ }\textbf {\bibinfo
  {volume} {92}},\ \bibinfo {pages} {035005} (\bibinfo {year} {2015})},\
  \Eprint {https://arxiv.org/abs/1505.03403} {arXiv:1505.03403 [hep-ph]}
  \BibitemShut {NoStop}%
\bibitem [{\citenamefont {Fonseca}(2021)}]{Fonseca:2020vke}%
  \BibitemOpen
  \bibfield  {author} {\bibinfo {author} {\bibfnamefont {R.~M.}\ \bibnamefont
  {Fonseca}},\ }\bibfield  {title} {\bibinfo {title} {{GroupMath: A Mathematica
  package for group theory calculations}},\ }\href
  {https://doi.org/10.1016/j.cpc.2021.108085} {\bibfield  {journal} {\bibinfo
  {journal} {Comput. Phys. Commun.}\ }\textbf {\bibinfo {volume} {267}},\
  \bibinfo {pages} {108085} (\bibinfo {year} {2021})},\ \Eprint
  {https://arxiv.org/abs/2011.01764} {arXiv:2011.01764 [hep-th]} \BibitemShut
  {NoStop}%
\bibitem [{\citenamefont {Feger}\ \emph {et~al.}(2020)\citenamefont {Feger},
  \citenamefont {Kephart},\ and\ \citenamefont {Saskowski}}]{Feger:2019tvk}%
  \BibitemOpen
  \bibfield  {author} {\bibinfo {author} {\bibfnamefont {R.}~\bibnamefont
  {Feger}}, \bibinfo {author} {\bibfnamefont {T.~W.}\ \bibnamefont {Kephart}},\
  and\ \bibinfo {author} {\bibfnamefont {R.~J.}\ \bibnamefont {Saskowski}},\
  }\bibfield  {title} {\bibinfo {title} {{LieART 2.0 \textendash{} A
  Mathematica application for Lie Algebras and Representation Theory}},\ }\href
  {https://doi.org/10.1016/j.cpc.2020.107490} {\bibfield  {journal} {\bibinfo
  {journal} {Comput. Phys. Commun.}\ }\textbf {\bibinfo {volume} {257}},\
  \bibinfo {pages} {107490} (\bibinfo {year} {2020})},\ \Eprint
  {https://arxiv.org/abs/1912.10969} {arXiv:1912.10969 [hep-th]} \BibitemShut
  {NoStop}%
\bibitem [{\citenamefont {Fritzsch}\ and\ \citenamefont
  {Minkowski}(1975)}]{Fritzsch:1974nn}%
  \BibitemOpen
  \bibfield  {author} {\bibinfo {author} {\bibfnamefont {H.}~\bibnamefont
  {Fritzsch}}\ and\ \bibinfo {author} {\bibfnamefont {P.}~\bibnamefont
  {Minkowski}},\ }\bibfield  {title} {\bibinfo {title} {{Unified Interactions
  of Leptons and Hadrons}},\ }\href
  {https://doi.org/10.1016/0003-4916(75)90211-0} {\bibfield  {journal}
  {\bibinfo  {journal} {Annals Phys.}\ }\textbf {\bibinfo {volume} {93}},\
  \bibinfo {pages} {193} (\bibinfo {year} {1975})}\BibitemShut {NoStop}%
\bibitem [{\citenamefont {Georgi}\ and\ \citenamefont
  {Glashow}(1974)}]{Georgi:1974sy}%
  \BibitemOpen
  \bibfield  {author} {\bibinfo {author} {\bibfnamefont {H.}~\bibnamefont
  {Georgi}}\ and\ \bibinfo {author} {\bibfnamefont {S.~L.}\ \bibnamefont
  {Glashow}},\ }\bibfield  {title} {\bibinfo {title} {{Unity of All Elementary
  Particle Forces}},\ }\href {https://doi.org/10.1103/PhysRevLett.32.438}
  {\bibfield  {journal} {\bibinfo  {journal} {Phys. Rev. Lett.}\ }\textbf
  {\bibinfo {volume} {32}},\ \bibinfo {pages} {438} (\bibinfo {year}
  {1974})}\BibitemShut {NoStop}%
\bibitem [{\citenamefont {Pati}\ and\ \citenamefont
  {Salam}(1974)}]{Pati:1974yy}%
  \BibitemOpen
  \bibfield  {author} {\bibinfo {author} {\bibfnamefont {J.~C.}\ \bibnamefont
  {Pati}}\ and\ \bibinfo {author} {\bibfnamefont {A.}~\bibnamefont {Salam}},\
  }\bibfield  {title} {\bibinfo {title} {{Lepton Number as the Fourth Color}},\
  }\href {https://doi.org/10.1103/PhysRevD.10.275} {\bibfield  {journal}
  {\bibinfo  {journal} {Phys. Rev. D}\ }\textbf {\bibinfo {volume} {10}},\
  \bibinfo {pages} {275} (\bibinfo {year} {1974})},\ \bibinfo {note} {[Erratum:
  Phys.Rev.D 11, 703--703 (1975)]}\BibitemShut {NoStop}%
\bibitem [{\citenamefont {Koren}(2022)}]{Koren:2022bam}%
  \BibitemOpen
  \bibfield  {author} {\bibinfo {author} {\bibfnamefont {S.}~\bibnamefont
  {Koren}},\ }\bibfield  {title} {\bibinfo {title} {{A Note on Proton Stability
  in the Standard Model}},\ }\href {https://doi.org/10.3390/universe8060308}
  {\bibfield  {journal} {\bibinfo  {journal} {Universe}\ }\textbf {\bibinfo
  {volume} {8}},\ \bibinfo {pages} {308} (\bibinfo {year} {2022})},\ \Eprint
  {https://arxiv.org/abs/2204.01741} {arXiv:2204.01741 [hep-ph]} \BibitemShut
  {NoStop}%
\bibitem [{\citenamefont {Wang}\ \emph
  {et~al.}(2022{\natexlab{b}})\citenamefont {Wang}, \citenamefont {Wan},\ and\
  \citenamefont {You}}]{Wang:2022eag}%
  \BibitemOpen
  \bibfield  {author} {\bibinfo {author} {\bibfnamefont {J.}~\bibnamefont
  {Wang}}, \bibinfo {author} {\bibfnamefont {Z.}~\bibnamefont {Wan}},\ and\
  \bibinfo {author} {\bibfnamefont {Y.-Z.}\ \bibnamefont {You}},\ }\bibfield
  {title} {\bibinfo {title} {{Proton stability: From the standard model to
  beyond grand unification}},\ }\href
  {https://doi.org/10.1103/PhysRevD.106.025016} {\bibfield  {journal} {\bibinfo
   {journal} {Phys. Rev. D}\ }\textbf {\bibinfo {volume} {106}},\ \bibinfo
  {pages} {025016} (\bibinfo {year} {2022}{\natexlab{b}})},\ \Eprint
  {https://arxiv.org/abs/2204.08393} {arXiv:2204.08393 [hep-ph]} \BibitemShut
  {NoStop}%
\bibitem [{\citenamefont {Allanach}\ \emph {et~al.}(2021)\citenamefont
  {Allanach}, \citenamefont {Gripaios},\ and\ \citenamefont
  {Tooby-Smith}}]{Allanach:2021bfe}%
  \BibitemOpen
  \bibfield  {author} {\bibinfo {author} {\bibfnamefont {B.~C.}\ \bibnamefont
  {Allanach}}, \bibinfo {author} {\bibfnamefont {B.}~\bibnamefont {Gripaios}},\
  and\ \bibinfo {author} {\bibfnamefont {J.}~\bibnamefont {Tooby-Smith}},\
  }\bibfield  {title} {\bibinfo {title} {{Semisimple extensions of the Standard
  Model gauge algebra}},\ }\href {https://doi.org/10.1103/PhysRevD.104.035035}
  {\bibfield  {journal} {\bibinfo  {journal} {Phys. Rev. D}\ }\textbf {\bibinfo
  {volume} {104}},\ \bibinfo {pages} {035035} (\bibinfo {year} {2021})},\
  \bibinfo {note} {[Erratum: Phys.Rev.D 106, 019901 (2022)]},\ \Eprint
  {https://arxiv.org/abs/2104.14555} {arXiv:2104.14555 [hep-th]} \BibitemShut
  {NoStop}%
\bibitem [{\citenamefont {Babu}\ \emph {et~al.}(1986)\citenamefont {Babu},
  \citenamefont {He},\ and\ \citenamefont {Pakvasa}}]{Babu:1985gi}%
  \BibitemOpen
  \bibfield  {author} {\bibinfo {author} {\bibfnamefont {K.~S.}\ \bibnamefont
  {Babu}}, \bibinfo {author} {\bibfnamefont {X.-G.}\ \bibnamefont {He}},\ and\
  \bibinfo {author} {\bibfnamefont {S.}~\bibnamefont {Pakvasa}},\ }\bibfield
  {title} {\bibinfo {title} {{Neutrino Masses and Proton Decay Modes in SU(3) X
  SU(3) X SU(3) Trinification}},\ }\href
  {https://doi.org/10.1103/PhysRevD.33.763} {\bibfield  {journal} {\bibinfo
  {journal} {Phys. Rev. D}\ }\textbf {\bibinfo {volume} {33}},\ \bibinfo
  {pages} {763} (\bibinfo {year} {1986})}\BibitemShut {NoStop}%
\bibitem [{\citenamefont {Willenbrock}(2003)}]{Willenbrock:2003ca}%
  \BibitemOpen
  \bibfield  {author} {\bibinfo {author} {\bibfnamefont {S.}~\bibnamefont
  {Willenbrock}},\ }\bibfield  {title} {\bibinfo {title} {{Triplicated
  trinification}},\ }\href {https://doi.org/10.1016/S0370-2693(03)00419-2}
  {\bibfield  {journal} {\bibinfo  {journal} {Phys. Lett. B}\ }\textbf
  {\bibinfo {volume} {561}},\ \bibinfo {pages} {130} (\bibinfo {year}
  {2003})},\ \Eprint {https://arxiv.org/abs/hep-ph/0302168}
  {arXiv:hep-ph/0302168} \BibitemShut {NoStop}%
\bibitem [{\citenamefont {Sayre}\ \emph {et~al.}(2006)\citenamefont {Sayre},
  \citenamefont {Wiesenfeldt},\ and\ \citenamefont
  {Willenbrock}}]{Sayre:2006ma}%
  \BibitemOpen
  \bibfield  {author} {\bibinfo {author} {\bibfnamefont {J.}~\bibnamefont
  {Sayre}}, \bibinfo {author} {\bibfnamefont {S.}~\bibnamefont {Wiesenfeldt}},\
  and\ \bibinfo {author} {\bibfnamefont {S.}~\bibnamefont {Willenbrock}},\
  }\bibfield  {title} {\bibinfo {title} {{Minimal trinification}},\ }\href
  {https://doi.org/10.1103/PhysRevD.73.035013} {\bibfield  {journal} {\bibinfo
  {journal} {Phys. Rev. D}\ }\textbf {\bibinfo {volume} {73}},\ \bibinfo
  {pages} {035013} (\bibinfo {year} {2006})},\ \Eprint
  {https://arxiv.org/abs/hep-ph/0601040} {arXiv:hep-ph/0601040} \BibitemShut
  {NoStop}%
\bibitem [{\citenamefont {Pelaggi}\ \emph {et~al.}(2015)\citenamefont
  {Pelaggi}, \citenamefont {Strumia},\ and\ \citenamefont
  {Vignali}}]{Pelaggi:2015kna}%
  \BibitemOpen
  \bibfield  {author} {\bibinfo {author} {\bibfnamefont {G.~M.}\ \bibnamefont
  {Pelaggi}}, \bibinfo {author} {\bibfnamefont {A.}~\bibnamefont {Strumia}},\
  and\ \bibinfo {author} {\bibfnamefont {S.}~\bibnamefont {Vignali}},\
  }\bibfield  {title} {\bibinfo {title} {{Totally asymptotically free
  trinification}},\ }\href {https://doi.org/10.1007/JHEP08(2015)130} {\bibfield
   {journal} {\bibinfo  {journal} {JHEP}\ }\textbf {\bibinfo {volume} {08}},\
  \bibinfo {pages} {130}},\ \Eprint {https://arxiv.org/abs/1507.06848}
  {arXiv:1507.06848 [hep-ph]} \BibitemShut {NoStop}%
\bibitem [{\citenamefont {Camargo-Molina}\ \emph {et~al.}(2016)\citenamefont
  {Camargo-Molina}, \citenamefont {Morais}, \citenamefont {Pasechnik},\ and\
  \citenamefont {Wess\'en}}]{Camargo-Molina:2016bwm}%
  \BibitemOpen
  \bibfield  {author} {\bibinfo {author} {\bibfnamefont {J.~E.}\ \bibnamefont
  {Camargo-Molina}}, \bibinfo {author} {\bibfnamefont {A.~P.}\ \bibnamefont
  {Morais}}, \bibinfo {author} {\bibfnamefont {R.}~\bibnamefont {Pasechnik}},\
  and\ \bibinfo {author} {\bibfnamefont {J.}~\bibnamefont {Wess\'en}},\
  }\bibfield  {title} {\bibinfo {title} {{On a radiative origin of the Standard
  Model from Trinification}},\ }\href {https://doi.org/10.1007/JHEP09(2016)129}
  {\bibfield  {journal} {\bibinfo  {journal} {JHEP}\ }\textbf {\bibinfo
  {volume} {09}},\ \bibinfo {pages} {129}},\ \Eprint
  {https://arxiv.org/abs/1606.03492} {arXiv:1606.03492 [hep-ph]} \BibitemShut
  {NoStop}%
\bibitem [{\citenamefont {Camargo-Molina}\ \emph {et~al.}(2019)\citenamefont
  {Camargo-Molina}, \citenamefont {Morais}, \citenamefont {Ordell},
  \citenamefont {Pasechnik},\ and\ \citenamefont
  {Wess\'en}}]{Camargo-Molina:2017kxd}%
  \BibitemOpen
  \bibfield  {author} {\bibinfo {author} {\bibfnamefont {J.~E.}\ \bibnamefont
  {Camargo-Molina}}, \bibinfo {author} {\bibfnamefont {A.~P.}\ \bibnamefont
  {Morais}}, \bibinfo {author} {\bibfnamefont {A.}~\bibnamefont {Ordell}},
  \bibinfo {author} {\bibfnamefont {R.}~\bibnamefont {Pasechnik}},\ and\
  \bibinfo {author} {\bibfnamefont {J.}~\bibnamefont {Wess\'en}},\ }\bibfield
  {title} {\bibinfo {title} {{Scale hierarchies, symmetry breaking and particle
  spectra in SU(3)-family extended SUSY trinification}},\ }\href
  {https://doi.org/10.1103/PhysRevD.99.035041} {\bibfield  {journal} {\bibinfo
  {journal} {Phys. Rev. D}\ }\textbf {\bibinfo {volume} {99}},\ \bibinfo
  {pages} {035041} (\bibinfo {year} {2019})},\ \Eprint
  {https://arxiv.org/abs/1711.05199} {arXiv:1711.05199 [hep-ph]} \BibitemShut
  {NoStop}%
\bibitem [{\citenamefont {Wang}\ \emph {et~al.}(2019)\citenamefont {Wang},
  \citenamefont {Al~Balushi}, \citenamefont {Mann},\ and\ \citenamefont
  {Jiang}}]{Wang:2018yer}%
  \BibitemOpen
  \bibfield  {author} {\bibinfo {author} {\bibfnamefont {Z.-W.}\ \bibnamefont
  {Wang}}, \bibinfo {author} {\bibfnamefont {A.}~\bibnamefont {Al~Balushi}},
  \bibinfo {author} {\bibfnamefont {R.}~\bibnamefont {Mann}},\ and\ \bibinfo
  {author} {\bibfnamefont {H.-M.}\ \bibnamefont {Jiang}},\ }\bibfield  {title}
  {\bibinfo {title} {{Safe Trinification}},\ }\href
  {https://doi.org/10.1103/PhysRevD.99.115017} {\bibfield  {journal} {\bibinfo
  {journal} {Phys. Rev. D}\ }\textbf {\bibinfo {volume} {99}},\ \bibinfo
  {pages} {115017} (\bibinfo {year} {2019})},\ \Eprint
  {https://arxiv.org/abs/1812.11085} {arXiv:1812.11085 [hep-ph]} \BibitemShut
  {NoStop}%
\bibitem [{\citenamefont {Barr}(1982)}]{Barr:1981qv}%
  \BibitemOpen
  \bibfield  {author} {\bibinfo {author} {\bibfnamefont {S.~M.}\ \bibnamefont
  {Barr}},\ }\bibfield  {title} {\bibinfo {title} {{A New Symmetry Breaking
  Pattern for SO(10) and Proton Decay}},\ }\href
  {https://doi.org/10.1016/0370-2693(82)90966-2} {\bibfield  {journal}
  {\bibinfo  {journal} {Phys. Lett. B}\ }\textbf {\bibinfo {volume} {112}},\
  \bibinfo {pages} {219} (\bibinfo {year} {1982})}\BibitemShut {NoStop}%
\bibitem [{\citenamefont {Barr}(1989)}]{Barr:1988yj}%
  \BibitemOpen
  \bibfield  {author} {\bibinfo {author} {\bibfnamefont {S.~M.}\ \bibnamefont
  {Barr}},\ }\bibfield  {title} {\bibinfo {title} {{Some Comments on Flipped
  SU(5) X U(1) and Flipped Unification in General}},\ }\href
  {https://doi.org/10.1103/PhysRevD.40.2457} {\bibfield  {journal} {\bibinfo
  {journal} {Phys. Rev. D}\ }\textbf {\bibinfo {volume} {40}},\ \bibinfo
  {pages} {2457} (\bibinfo {year} {1989})}\BibitemShut {NoStop}%
\bibitem [{\citenamefont {Derendinger}\ \emph {et~al.}(1984)\citenamefont
  {Derendinger}, \citenamefont {Kim},\ and\ \citenamefont
  {Nanopoulos}}]{Derendinger:1983aj}%
  \BibitemOpen
  \bibfield  {author} {\bibinfo {author} {\bibfnamefont {J.~P.}\ \bibnamefont
  {Derendinger}}, \bibinfo {author} {\bibfnamefont {J.~E.}\ \bibnamefont
  {Kim}},\ and\ \bibinfo {author} {\bibfnamefont {D.~V.}\ \bibnamefont
  {Nanopoulos}},\ }\bibfield  {title} {\bibinfo {title} {{Anti-SU(5)}},\ }\href
  {https://doi.org/10.1016/0370-2693(84)91238-3} {\bibfield  {journal}
  {\bibinfo  {journal} {Phys. Lett. B}\ }\textbf {\bibinfo {volume} {139}},\
  \bibinfo {pages} {170} (\bibinfo {year} {1984})}\BibitemShut {NoStop}%
\end{thebibliography}%
\let\addcontentsline\oldaddcontentsline

\end{document}